\documentclass[12pt]{article}
\usepackage{newtxtext}
\usepackage{amssymb}
\usepackage{url}
\usepackage{float}
\usepackage[top=20mm, bottom=25mm, left=30mm, right=25mm]{geometry} 
\usepackage[labelformat=simple]{subcaption}
\usepackage{amsmath}

\usepackage{xcolor}
\usepackage[T1]{fontenc}
\usepackage{graphicx}
\usepackage{authblk}

\title{Generalised envelope spectrum-based signal-to-noise objectives: Formulation, optimisation and application for gear fault detection under time-varying speed conditions }

\author[1]{Stephan Schmidt\thanks{Corresponding author. Email: \texttt{stephan.schmidt@up.ac.za}}}
\author[1]{Daniel N. Wilke\thanks{Email: \texttt{nico.wilke@up.ac.za}}}
\author[2,3]{Konstantinos C. Gryllias\thanks{Email: \texttt{konstantinos.gryllias@kuleuven.be}}}
\affil[1]{\small Centre for Asset Integrity Management, Department of Mechanical and Aeronautical Engineering, University of Pretoria, Pretoria, South Africa.}
\affil[2]{\small Department of Mechanical Engineering, KU Leuven, Celestijnenlaan 300, 3001 Heverlee, Belgium.}
\affil[3]{\small Flanders Make Belgium.}

\begin{document}

\date{}

\maketitle

\begin{abstract}
In vibration-based condition monitoring, optimal filter design improves fault detection by enhancing weak fault signatures within vibration signals. This process involves optimising a derived objective function from a defined objective. The objectives are often based on proxy health indicators to determine the filter's parameters. However, these indicators can be compromised by irrelevant extraneous signal components and fluctuating operational conditions, affecting the filter's efficacy. Fault detection primarily uses the fault component's prominence in the squared envelope spectrum, quantified by a squared envelope spectrum-based signal-to-noise ratio. New optimal filter objective functions are derived from the proposed generalised envelope spectrum-based signal-to-noise objective for machines operating under variable speed conditions. Instead of optimising proxy health indicators, the optimal filter coefficients of the formulation directly maximise the squared envelope spectrum-based signal-to-noise ratio over targeted frequency bands using standard gradient-based optimisers. Four derived objective functions from the proposed objective effectively outperform five prominent methods in tests on three experimental datasets. 
\end{abstract}

%
%
%
%
%
%


\section{Introduction}

Vibration-based condition monitoring is frequently used to monitor critical rotating assets such as gearboxes. The diagnostic information in the collected vibration signals is often masked by signal components generated by high-energy interactions between healthy rotating components \cite{borghesani2013application}, the operating conditions \cite{schmidt2020normalisation}, and the operating environment \cite{hebda2022infogram}. These signal components impede the performance of conventional vibration-based condition monitoring methods and can result in delayed detection and unexpected breakdowns due to false negatives. 

Automatic filter design methods can be used to construct filters that enhance the fault signatures and attenuate the extraneous components \cite{smith2019optimal,miao2022review}. Informative frequency band methods often use a specific filter parametrisation to formulate the filter design problem as a two-dimensional optimisation problem where the lower and upper bounds of the bandpass filters are the design variables. However, the exact fault signatures are unknown. Therefore, proxy objectives for damage such as the spectral negentropy \cite{antoni2016infogram}, the $L_2$/$L_1$-norm \cite{wang2018spectral}, the ratio of the cyclic content \cite{borghesani2014relationship}, the Indicator of second-order CycloStationarity (ICS2) \cite{smith2019optimal} and envelope spectrum-based signal-to-noise objectives (e.g., \cite{mauricio2020improved,schmidt2021informative}) are used to derive objective functions to determine optimal filter parameters. 

The optimal filter design problem can also be formulated to determine all filter coefficients by optimising an objective function, derived from proxy objectives for damage \cite{miao2022review}. This is referred to as filter coefficient optimisation, and it provides the advantage that the filter's frequency response has more flexibility to extract the underlying fault information. Time domain-based methods such as the minimum entropy deconvolution \cite{sawalhi2007enhancement} and the generalised $L_p$/$L_q$ ratio \cite{jia2018sparse} have been used for fault signature enhancement. Since the amplitude modulation due to the damaged gear and bearing components manifest at sparse cyclic orders in the envelope spectrum, sparsity proxy measures such as the spectral negentropy \cite{peeters2020blind}, the Gini index \cite{miao2022practical}, the $L_2$/$L_1$-norm \cite{peeters2020blind}, the generalised $L_p$/$L_q$-norm \cite{he2021extracting}, and the modified smoothness index \cite{chen2023squared} are examples of metrics used to formulate objective functions for optimal filter design to detect damage. 
Recently, formulations include non-linear generalised Rayleigh quotients to conduct blind deconvolution \cite{buzzoni2018blind,soave2022blind,peeters2020blind}.

Instead of using metrics that measure a statistic of the entire time domain signal or entire envelope spectrum, metrics can be defined as those that target the properties of specific components. These metrics are referred to as targeted metrics \cite{smith2019optimal} and can be more robust to extraneous impulsive components \cite{meng2023maximum}. McDonald and Zhao \cite{mcdonald2017multipoint} proposed a Multi-point Optimal Minimum Entropy Deconvolution Adjusted (MOMEDA) method that finds the optimal filter coefficients in closed form by maximising the multi-D-norm of the signal. Buzzoni et al. \cite{buzzoni2018blind} proposed the maximum second-order CYClostationarity Blind Deconvolution (CYCBD), which uses a targeted objective function based on the ICS2. Soave et al. \cite{soave2022blind} replaced the Fourier transform of the CYCBD with a Fourier-Bessel expansion for improved performance. The CYCBD targets the theoretical cyclic orders of the fault. Chen et al. \cite{chen2020blind} investigated different periodicity detection techniques to circumvent the need to specify the targeted cyclic order a priori in the CYCBD. Zhang et al. \cite{zhang2021adaptive} proposed an Adaptive maximum second-order CYClostationarity Blind Deconvolution (ACYCBD) method, which uses the envelope harmonic product spectrum to determine the actual cyclic order of the damaged component before calculating the ICS2 of the filtered signal. Wang et al. \cite{wang2022bearing} proposed another ACYCBD method, which determines the targeted cyclic frequency and the filter length for the CYCBD.

Other targeted metrics have recently been proposed for optimal filter design formulations. Liang et al. \cite{liang2021maximum} proposed maximising the average kurtosis of each fault period.  Meng et al. \cite{meng2023maximum} proposed the Maximum cyclic Gini index deconvolution, which calculates the Gini index of specifically targeted components and overcomes the Gini index's sensitivity to extraneous components. Peng et al. \cite{peng2023use} proposed maximising the generalised Gaussian cyclostationarity indicator to determine the filter's coefficients. Zhou et al. \cite{zhou2022blind} proposed the Maximum Squared Envelope Spectrum Harmonic-to-Interference Ratio Deconvolution (MSESHIRD) method, which maximises the ratio of the sum of the amplitudes in specific frequency bands in the Squared Envelope Spectrum (SES) to the difference between the sum of all the amplitudes in the SES and the sum of the amplitudes in specific frequency bands in the SES. 

Various optimisation approaches are available to solve filter coefficient optimisation problems. Lee and Nandi \cite{lee1998blind} obtained stationary points to search for optimal filter coefficients that maximised the filtered signal's higher-order statistics. This approach has been used for extracting impacting signals \cite{lee2000extraction} and also for enhancing bearing damage \cite{sawalhi2007enhancement}. Generalised Rayleigh quotients are generally solved using an iterative generalised eigenvalue decomposition \cite{buzzoni2018blind,soave2022blind,peeters2020blind}. He et al. \cite{he2021extracting}, and Zhou et al. \cite{zhou2022blind} used analytical gradients with gradient-based minimisers to find the optimal filter coefficients. Fang et al. \cite{fang2021blind} proposed a general backward automatic differentiation method for a predefined objective function to compute gradients that are then used to find sets of optimal filter coefficients.

Many critical power generation and mining assets are exposed to time-varying speed conditions \cite{zimroz2014diagnostics}. Time-varying speed conditions often impede the performance of methods developed for constant speed conditions, however, few research works have considered the impact of varying speed conditions. Liang et al. \cite{liang2021maximum} applied the maximum average kurtosis under varying speed conditions. Buzzoni et al. \cite{buzzoni2018blind} proposed and investigated an extension of the CYCBD for time-varying speed conditions utilising the Velocity Synchronous Discrete Fourier Transform (VS-DFT) proposed by Borghesani et al. \cite{borghesani2014velocity}. Zhou et al. \cite{zhou2022blind} used a cyclic order band to accommodate speed fluctuations.

The rotating machinery fault detection literature has demonstrated the potential of envelope spectrum-based signal-to-noise metrics (e.g., \cite{mauricio2020improved,schmidt2021informative}). Furthermore, the prominence of the amplitudes in the envelope spectrum is often used to diagnose the machine, e.g., if the ball pass outer race component is more prominent than the noise floor in the envelope spectrum, there is potential outer race damage in the system \cite{randall2011rolling}. However, according to our knowledge, the envelope spectrum-based signal-to-noise ratios have yet to be used for filter coefficient optimisation under time-varying speed conditions.  Hence, a generalised objective is proposed that makes it possible to maximise the envelope spectrum-based signal-to-noise ratio as well as other metrics, such as the ICS2, under time-varying speed conditions. The filtered signal obtained by maximising the proposed objective is compared against other objectives under time-varying conditions. New objectives can be derived from the generalised objective and existing objectives estimated, such as the ICS2 and the MSESHIRD, for time-varying speed applications. In summary, the following primary contributions are made:
\begin{itemize}
	\item A Generalised Envelope spectrum-based Signal-to-Noise (GES2N) objective is proposed that can be used to estimate the ICS2 and the MSESHIRD and to develop squared envelope-based signal-to-noise optimal filter formulations for filter coefficient optimisation.
	\item This proposed generalised targeted objective is specifically derived for time-varying speed conditions.
	\item Different objectives are derived and compared against existing methods for gear damage detection under time-varying speed conditions.
\end{itemize}

This work establishes terminology regarding metrics, objectives, and objective functions. Firstly, the objective function to be optimised can differ from the objective being optimised. The objective represents a task's overarching goal, aim or idea, e.g. maximising the signal-to-noise ratio. Secondly, the objective function is the function that is {optimised} to realise the objective. This function could be the original or some transformed objective that enhances the optimisability of the objective. Objective functions with underlying characteristics should be appropriately paired with optimisers' fundamental assumptions of objective functions \cite{snyman2005practical}. Lastly, objectives and objective functions merely {evaluated} on data can be used as metrics for fault detection compared to thresholds for automated fault detection.

The paper is structured as follows: Section \ref{sec:Method:main} presents an overview of the definition and optimisation of the proposed objective, whereafter, the method is investigated on three experimental datasets in Section \ref{sec:ExpData:main}. Finally, the work concludes in Section \ref{sec:Conclusion:main} with recommendations for future work. Appendix~\ref{app:GitLabCode} contains a link to a GitLab page with the Python code of the proposed GES2N objective. Appendix~\ref{app:DerivationGradient} contains the gradient of the objective function.

\section{Proposed optimal filter design method}
\label{sec:Method:main}
\begin{figure}[h]
	\centering
	\includegraphics[width=0.987\linewidth]{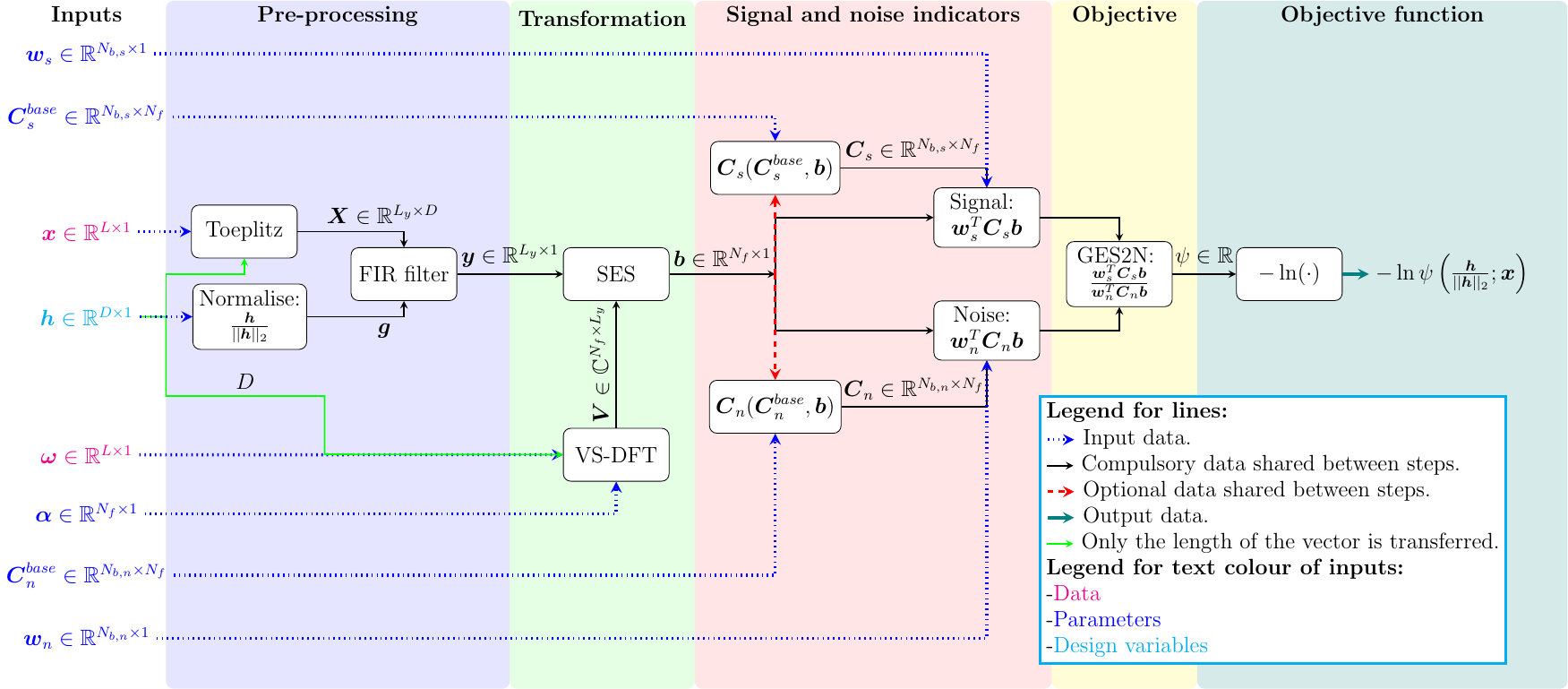}
	\caption{An overview of the evaluation of the proposed Generalised Envelope Spectrum-Based Signal-to-Noise (GES2N) objective and associated objective function. Abbreviations: Finite Impulse Response (FIR); Squared Envelope Spectrum (SES); Velocity Synchronous Discrete Fourier Transform (VS-DFT). The natural logarithm is denoted $\ln$.}
	\label{fig:Method:OverviewOfMethod}
\end{figure}

Figure \ref{fig:Method:OverviewOfMethod} provides an overview of the evaluation procedure of the proposed targeted objective function, with a detailed overview and motivation of the steps presented in subsequent sections. 

The vibration signal $\boldsymbol{x}$ is filtered with $L_2$-normalised filter coefficients $\boldsymbol{g} = \boldsymbol{h}/||\boldsymbol{h}||_{2}$ to obtain the filtered signal $\boldsymbol{y}$, where $\boldsymbol{h}$ denotes the non-normalised filter coefficients. After that, the SES of the filtered signal $\boldsymbol{b}$ is calculated using the VS-DFT matrix $\boldsymbol{V}$, the rotational speed of a reference shaft $\boldsymbol{\omega}$ and the cyclic order vector $\boldsymbol{\alpha}$. Two indicators are calculated to form the proposed objective: an indicator of the signal (i.e., the information that needs to be maximised) and an indicator for the noise (i.e., the information that needs to be minimised or attenuated). The signal term in the numerator $\boldsymbol{w}_{s}^\textrm{T} \boldsymbol{C}_{s} \boldsymbol{b}$ and noise term in the denominator $\boldsymbol{w}_{n}^\textrm{T} \boldsymbol{C}_{n} \boldsymbol{b}$ have the same form, however, different cyclic band weight vectors $\boldsymbol{w}_i$ and different cyclic order weight matrices $\boldsymbol{C}_i$ are used with $i\in \left\{s,n\right\}$. The cyclic order weight matrices $\boldsymbol{C}_i$ are obtained by processing base matrices $\boldsymbol{C}_i^{base}$, where the general processing function is denoted $\boldsymbol{C}_i(\boldsymbol{C}_i^{base}, \boldsymbol{b})$, with $i \in \{s,n\}$. The SES can process the base matrices and, therefore, is included in the general processing function. Finally, the objective is defined as the ratio of the signal and noise terms, and the minimisation objective function is defined as the objective's negative logarithm.


In the next section, the finite impulse response filtering process is described.

\subsection{Finite Impulse Response (FIR) filtering}
\label{sec:Method:Filtering}
The measured vibration signal $\boldsymbol{x} = [x[0],x[1],\ldots,x[L-1]]^\textrm{T} \in \mathbb{R}^{L\times 1}$ is sampled at a constant frequency of $f_s$, such that the discrete time corresponding to $x[n]$ is given by $t[n] = n/f_s$, where $n$ is the time index. The angular speed vector of the reference shaft, measured in radians per second, is denoted $\boldsymbol{\omega} = [\omega[0],\omega[1],\ldots,\omega[L-1]]^\textrm{T} \in \mathbb{R}^{L\times 1}$ and the corresponding instantaneous angle of the shaft is denoted $\boldsymbol{\theta} = [\theta[0],\theta[1],\ldots,\theta[L-1]]^\textrm{T} \in \mathbb{R}^{L\times 1}$, with $\theta[0] = 0$ radians. Using the trapezoidal integration rule, the instantaneous angle can be estimated from the angular speed.

A filtered vibration signal $\boldsymbol{y} = \boldsymbol{x}\otimes\boldsymbol{g}, \text{ with } \boldsymbol{y} \in \mathbb{R}^{L_y \times 1}$, is obtained by convolving the measured signal $\boldsymbol{x}$ with the FIR filter's coefficients $\boldsymbol{g} \in \mathbb{R}^{D\times 1}$, with the filter's length denoted $D$ and $L_{y} = L - D -1$. The convolution process can be written as a matrix-vector product $\boldsymbol{y} = \boldsymbol{X} \boldsymbol{g}$ \cite{buzzoni2018blind}, where $\boldsymbol{X}\in \mathbb{R}^{L_{y} \times D}$ is defined as follows \cite{buzzoni2018blind}: 
\begin{equation}
	\boldsymbol{X} = \left[
	\begin{array}{c c c c}
		x[D-1] & x[D-2] & \ldots & x[0] \\
		x[D] & x[D-1] & \ldots & x[1] \\
		\vdots & \vdots & \ddots & \vdots \\
		x[L_y+D-2] & x[L_y-D-3] & \ldots & x[L_y-1] \\
	\end{array}
	\right].
\end{equation}
The aim is to find $\boldsymbol{g}$ such that the fault signatures are enhanced in the filtered signal $\boldsymbol{y}$ and the extraneous components are attenuated. Since gear and bearing damage often manifest as periodic modulation, it is useful to calculate objectives from the SES. 

\subsection{Squared Envelope Spectrum (SES)}
\label{sec:Method:SES}
The SES is often used in rotating machinery diagnostics (e.g., \cite{randall2011rolling,schmidt2021informative}. The order-domain SES, denoted $\boldsymbol{b} \in \mathbb{R}^{N_f \times 1}$, with the corresponding cyclic orders $\boldsymbol{\alpha} = [\alpha[0], \alpha[1], \ldots, \alpha[N_f-1]]^\textrm{T} \in \mathbb{R}^{N_{f}\times 1}$, is estimated with
\begin{equation}
	\boldsymbol{b} = (\boldsymbol{V} \left(
	\boldsymbol{y} \odot \boldsymbol{y}
	\right))^{\boldsymbol{*}} \odot (\boldsymbol{V} \left(
	\boldsymbol{y} \odot \boldsymbol{y}
	\right)),
\end{equation}
where ${}^{\boldsymbol{*}}$ denotes the element-wise conjugate operator, $\odot$ denotes the Hadamard product, and $\boldsymbol{V} \in \mathbb{C}^{N_f \times L_y}$ denotes the velocity synchronous Fourier matrix \cite{borghesani2014velocity}, with the number of cyclic orders denoted ${N_f}$. Each element of the velocity synchronous matrix is calculated with \cite{borghesani2014velocity,buzzoni2018blind}
\begin{equation}
	V[m,{s}] = \frac{1}{f_s \cdot \theta[L_y-1]} \cdot \omega[n] \cdot e^{-j \cdot \alpha[m] \cdot \theta[n]},
\end{equation}
with the rotational speed of the shaft at time increment $n$ denoted $\omega[n]$, the instantaneous angle of the shaft at time increment $n$ denoted $\theta[n]$, $j = \sqrt{-1}$, $m \in \{0,1,\ldots,N_{f}-1\}$, and $n \in \{0,1,\ldots,L_y-1\}$. The cyclic order resolution is fixed and equal to $\Delta \alpha = \frac{\alpha[N_f-1]}{N_f}$, with $\alpha[k] = k \cdot \Delta \alpha, \, \forall \, k \in \{0,1,\ldots,N_{f}-1\}$.

\begin{figure}[h]
	\centering
	\begin{subfigure}{0.49\linewidth}
		\caption{}
		\includegraphics[width=0.97\linewidth]{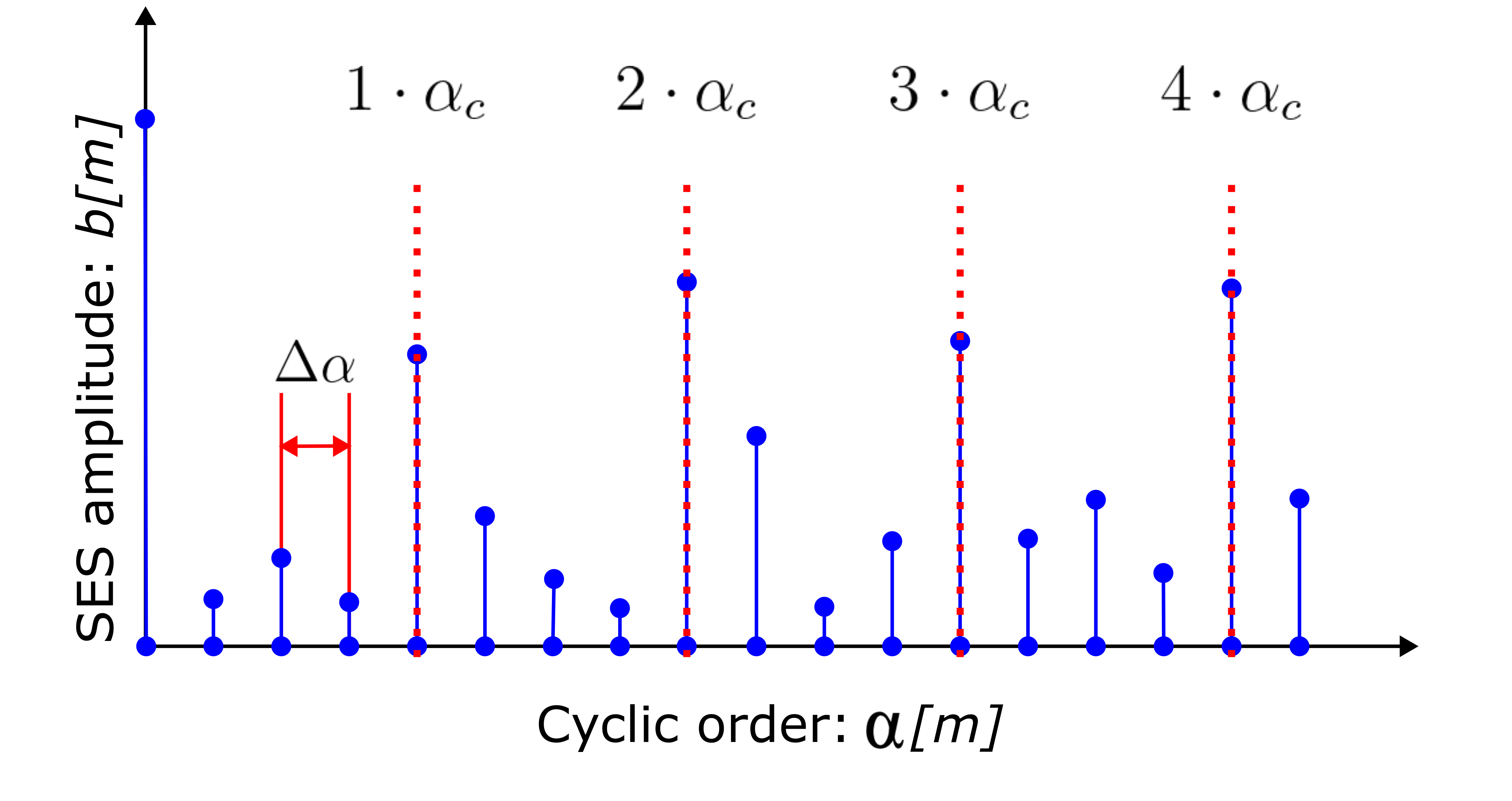}
		\label{fig:Method:SES_Example:A}		
	\end{subfigure}
	\begin{subfigure}{0.49\linewidth}
		\caption{}
		\includegraphics[width=0.97\linewidth]{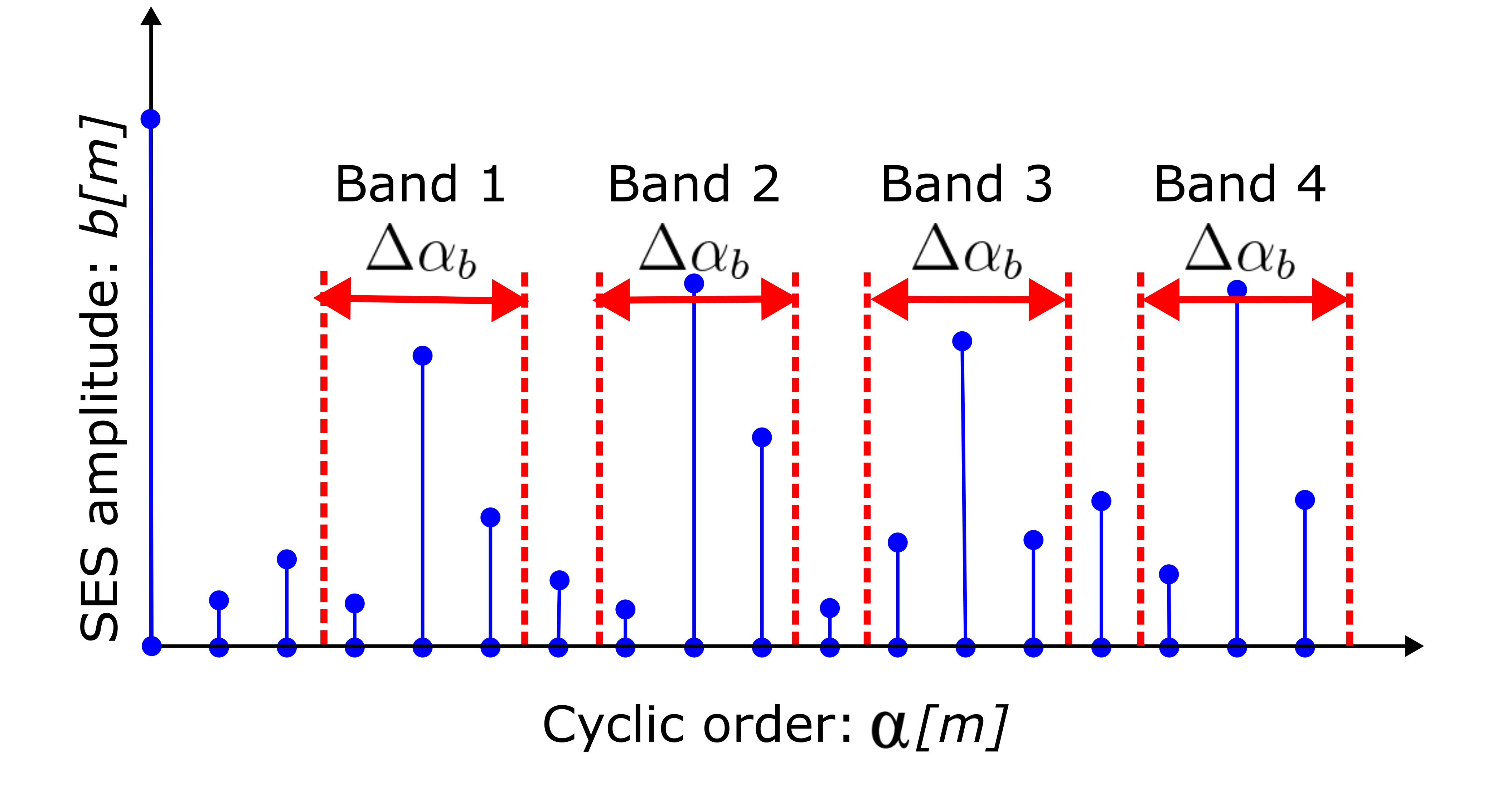}
		\label{fig:Method:SES_Example:B}
	\end{subfigure}	
	
	\caption{The SES amplitudes  $\boldsymbol{b}$ is presented against the corresponding cyclic orders $\boldsymbol{\alpha}$, with the amplitude corresponding to the cyclic order $\alpha[m]$ denoted $b[m]$. In (a), the cyclic order resolution $\Delta \alpha$ and the amplitudes of four targeted harmonics $k \cdot \alpha_{c}$ are shown, i.e., $N_h = 4$. In (b), four cyclic bands with a constant bandwidth $\Delta \alpha_{b}$ around the targeted harmonics $k \cdot \alpha_{c}$ are shown. }
	\label{fig:Method:SES_Example}
\end{figure}

A SES with a cyclic order resolution of $\Delta \alpha$ is shown in Figure \ref{fig:Method:SES_Example:A}, with the signal components associated with a cyclic order of $\alpha_c$ and its harmonics highlighted in Figure \ref{fig:Method:SES_Example:A}. Instead of monitoring a specific cyclic order (e.g., $\alpha_c$), monitoring and extracting metrics from bands around the theoretically targeted components is possible. An example of cyclic order bands with a constant bandwidth around the targeted cyclic orders $\alpha_c$ is shown in Figure \ref{fig:Method:SES_Example:B}. The bandwidth $\Delta \alpha_{b}$ is constant in this work but can be selected to be a function of the cyclic order, e.g., $\Delta \alpha_{b}(\alpha_{c})$.  The cyclic order bands accommodate slight deviations from the expected cyclic order. The next section defines the proposed objective for deriving specific objective functions for filter coefficient optimisation.

\subsection{Generalised Envelope spectrum-based Signal-to-Noise (GES2N) objective}
\label{sec:Method:GES2N}
The damaged bearing and gear signal components manifest at specific cyclic orders in the SES. This information has been utilised to define targeted filter coefficient optimisation methods such as the CYCBD and the MSESHIRD. Other objectives have been used for informative frequency band identification that also targets specific cyclic orders, such as the ratio of cyclic content \cite{borghesani2014relationship}, the ICS2gram \cite{smith2019optimal} (which uses the same objective as the CYCBD), the IESFOgram \cite{mauricio2020improved}, and the IFBI$_{\alpha}$gram \cite{schmidt2021informative}. These objectives are the ratios between two linear functions in the SES.  Hence, the following generalised objective is proposed\footnote{This representation is not unique, for example, we can write $\frac{\boldsymbol{w}_{s}^\textrm{T}\boldsymbol{C}_{s}\boldsymbol{b}}{\boldsymbol{w}_{n}^\textrm{T}\boldsymbol{C}_{n}\boldsymbol{b}} = \frac{\boldsymbol{c}_{s}^\textrm{T} \boldsymbol{b}}{\boldsymbol{c}_{n}^\textrm{T}\boldsymbol{b}}$ where $\boldsymbol{c}_i$ contains the weighting of the cyclic order components, i.e., $ \boldsymbol{c}_i = \boldsymbol{w}_{i}^\textrm{T}\boldsymbol{C}_i,\;\;i\in\{s,n\}$. However, the specific objective separates the cyclic order and cyclic band weightings. }:
\begin{equation}
	\label{eqn:Method:MainObjective}
	\psi\left(
	\boldsymbol{g}; \boldsymbol{x}, \boldsymbol{w}_{s}, \boldsymbol{C}_{s}, \boldsymbol{w}_{n}, \boldsymbol{C}_{n}
	\right) = \frac{\boldsymbol{w}_{s}^\textrm{T}\boldsymbol{C}_{s}\boldsymbol{b}}{\boldsymbol{w}_{n}^\textrm{T}\boldsymbol{C}_{n}\boldsymbol{b}},
\end{equation}
where $\psi: \mathbb{R}^{D\times 1} \mapsto \mathbb{R}$.
The objective in Equation \eqref{eqn:Method:MainObjective} makes it possible to define existing targeted methods such as the ICS2 and also signal-to-noise ratio measures in the SES, where the numerator contains the information that needs to be maximised (e.g., the fault signatures) and the denominator contains information that needs to be attenuated (e.g., the noise floor in the SES, the extraneous components). The objective in Equation \eqref{eqn:Method:MainObjective} is subsequently denoted $\psi\left(
\boldsymbol{g}; \boldsymbol{x}
\right)$ to simplify the notation.

\begin{figure}[h]
	\centering
	\begin{subfigure}{0.49\linewidth}
		\caption{}
		\includegraphics[width=0.97\linewidth]{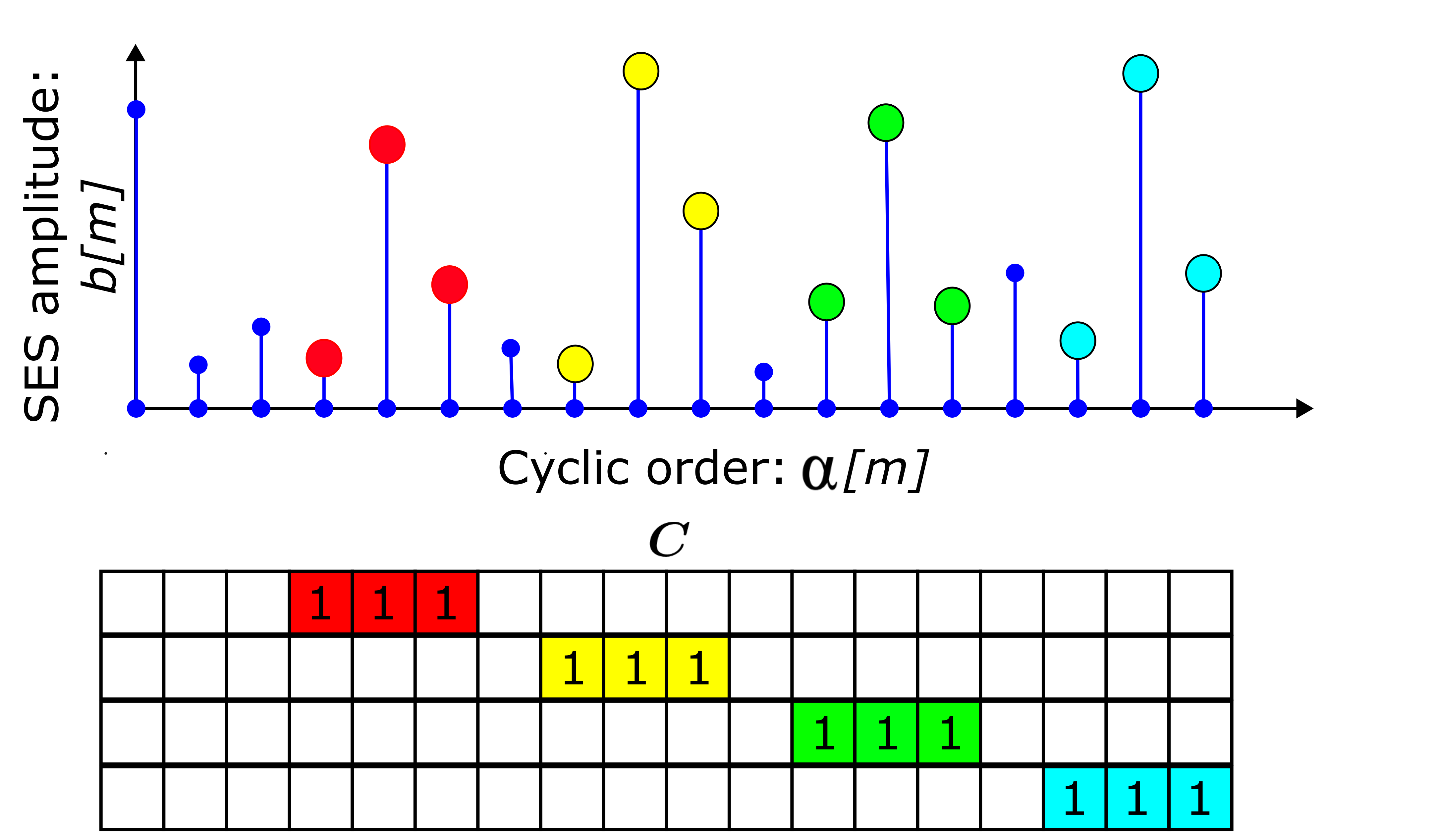}
		\label{fig:Method:WeightingMatrix_Sum}		
	\end{subfigure}
	\begin{subfigure}{0.49\linewidth}
		\caption{}
		\includegraphics[width=0.97\linewidth]{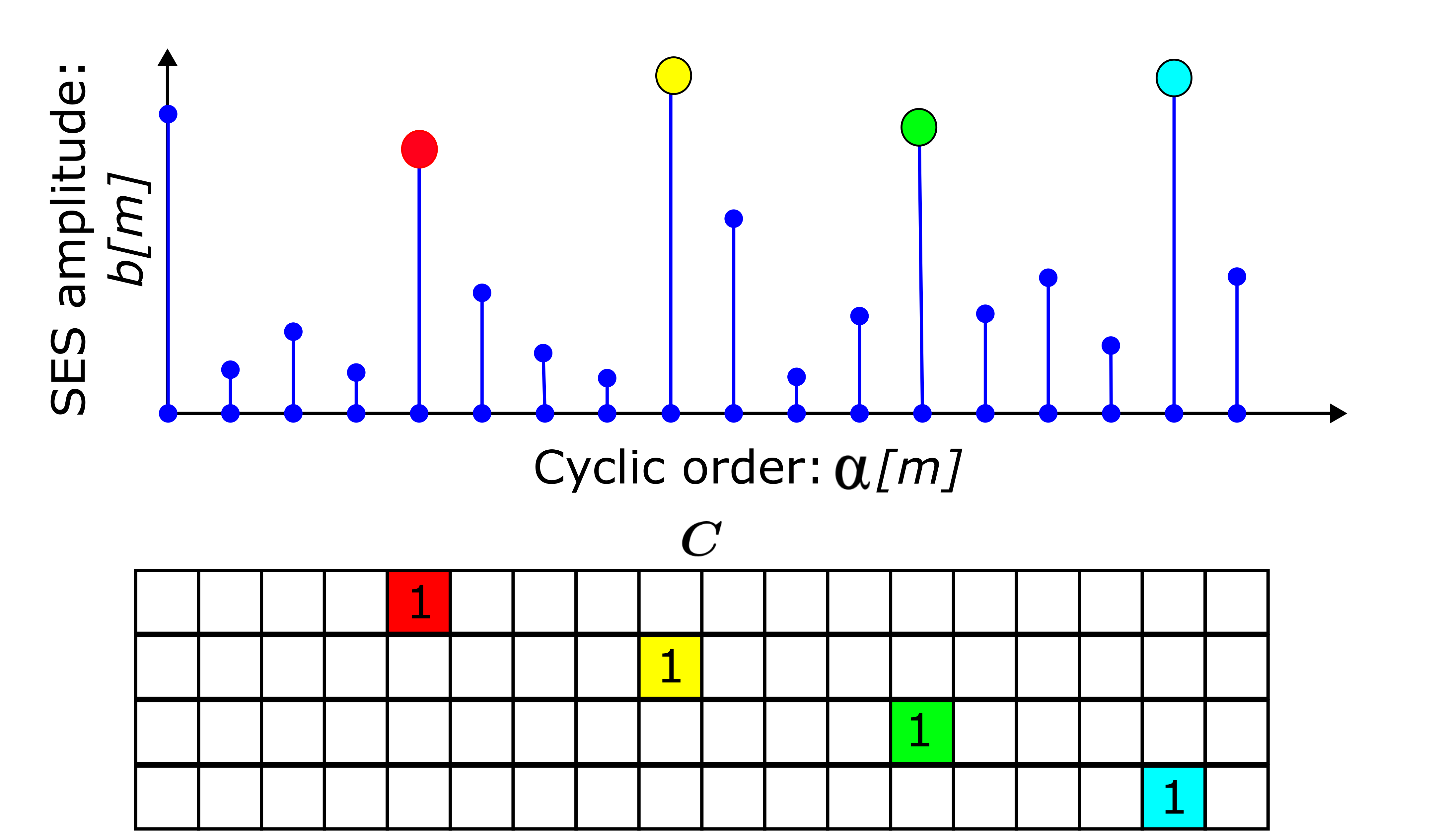}
		\label{fig:Method:WeightingMatrix_Max}
	\end{subfigure}
	\caption{Two examples of weighting matrices are shown for the Squared Envelope Spectrum (SES) in Figure \ref{fig:Method:SES_Example}(b). The SES has four bands $N_b = 4$. If the sum of the components in each band is calculated (e.g., like the MSESHIRD objective function), the cyclic order weighting matrix in Figure \ref{fig:Method:WeightingMatrix}(a) can be used. If the maximum of each band needs to be calculated, the cyclic order weighting matrix has the form shown in Figure \ref{fig:Method:WeightingMatrix}(b).  }
	\label{fig:Method:WeightingMatrix}
\end{figure}

The SES, denoted $\boldsymbol{b} \in \mathbb{R}^{N_f \times 1}$, of the filtered signal $\boldsymbol{y} = \boldsymbol{x} \otimes \boldsymbol{g}$ is assumed to contain $N_{b,s}$ potential cyclic order bands of interest in the numerator and $N_{b,d}$ potential cyclic order bands of interest in the denominator (refer to Figure \ref{fig:Method:SES_Example} for an overview of cyclic order bands). The cyclic order weighting matrices denoted $\boldsymbol{C}_{s} \in \mathbb{R}^{N_{b,s} \times N_f}$ and $\boldsymbol{C}_{n} \in \mathbb{R}^{N_{b,d} \times N_f}$ for the numerator and the denominator respectively, contain the weightings for each cyclic order $\alpha$ in each cyclic order band. Two examples of cyclic order weighting matrices are shown in Figure \ref{fig:Method:WeightingMatrix} for the SES in Figure \ref{fig:Method:SES_Example}. If the cyclic order weighting matrix in Figure \ref{fig:Method:WeightingMatrix}(a) is used to calculate $\boldsymbol{a} = \boldsymbol{C} \boldsymbol{b}$, then $a[m]$ provides the sum of the components in the $m$th cyclic order band. In contrast, if the cyclic order weighting matrix in Figure  \ref{fig:Method:WeightingMatrix}(b) is used to calculate $\boldsymbol{a} = \boldsymbol{C} \boldsymbol{b}$, then $a[m]$ contains the maximum amplitude in each cyclic order band.  The vectors $\boldsymbol{w}_{s} \in \mathbb{R}^{N_{b,s}\times 1}$ and $\boldsymbol{w}_{n} \in \mathbb{R}^{N_{b,d}\times 1}$ contain the weighting between the cyclic order bands in the numerator and the denominator respectively. In this work, $\boldsymbol{w}_{s} = \boldsymbol{1} \in \mathbb{R}^{N_{b,s} \times 1}$ and $\boldsymbol{w}_{n} = \boldsymbol{1} \in \mathbb{R}^{N_{b,d} \times 1}$, i.e., all cyclic order bands are weighted the same.

\subsection{Optimisation formulation}
\label{sec:Method: Optimisation}

The optimal filtering problem is formulated as the following unconstrained minimisation problem:
\begin{equation}
	\label{eqn:Method:Formulation2}
	\min_{\boldsymbol{h}}-\ln\psi\left(
	\frac{\boldsymbol{h}}{|| \boldsymbol{h}||_{2}}; \boldsymbol{x}
	\right),
\end{equation}
where $\boldsymbol{h}$ denotes the unconstrained filter coefficients and $\boldsymbol{g} = \frac{\boldsymbol{h}}{|| \boldsymbol{h}||_{2}} \in \mathbb{R}^{D \times 1}$ denotes the $L_2$-normalised filter coefficients, with $|| \boldsymbol{h}||_{2} = \sqrt{\boldsymbol{h}^\textrm{T} \boldsymbol{h}}$. The minimisation objective function is denoted $-\ln\psi\left(
	\boldsymbol{g}; \boldsymbol{x}
	\right)$, where $\psi\left(\boldsymbol{g}; \boldsymbol{x}
	\right)$ is defined in Equation \eqref{eqn:Method:MainObjective}. We found that the logarithm of the objective in Equation \eqref{eqn:Method:MainObjective} was easier to optimise than the objective itself. However, finding the best formulation is not in the scope of this work. The natural logarithm of the objective in Equation \eqref{eqn:Method:MainObjective}, evaluated with the normalised filter coefficients $\boldsymbol{g}$, is given by
\begin{equation}
	\label{eqn:Method:MainObjective_Log}
	\ln\psi\left(
	\boldsymbol{g}; \boldsymbol{x}
	\right) = \ln\left(
	\boldsymbol{w}_{s}^\textrm{T}\boldsymbol{C}_{s}\boldsymbol{b}
	\right) - \ln\left(
	\boldsymbol{w}_{n}^\textrm{T}\boldsymbol{C}_{n}\boldsymbol{b}
	\right),
\end{equation}
with the filtered signal obtained using $\boldsymbol{y} = \boldsymbol{X} \boldsymbol{g}  = \boldsymbol{X}\frac{\boldsymbol{h}}{|| \boldsymbol{h}||_{2}}$.

Gradient-based minimisers with analytical gradients are used as these methods are better suited than gradient-free minimisers and population-based solvers for high-dimensional problems. The gradient-based minimisers perform well for filter coefficient optimisation problems, and several well-established minimisers are available in the literature \cite{snyman2005practical} and programming language libraries (e.g., NLopt \cite{NLopt}, Julia's Optim \cite{mogensen2018optim} and Python's SciPy modules \cite{2020SciPy-NMeth}).  The gradient of Equation \eqref{eqn:Method:MainObjective_Log} is defined in Appendix~\ref{app:DerivationGradient}.

We found the Conjugate Gradient (CG) method, which is generally available in optimisation packages (e.g., Python's SciPy \cite{2020SciPy-NMeth}, Julia's Optim \cite{mogensen2018optim}), perform sufficiently well to solve the optimisation problem in Equation \eqref{eqn:Method:Formulation2} when using linear predictive coding \cite{buzzoni2018blind} to initialise the filter. Finding the best combination of formulations, solvers, and initialisation strategies is beyond the scope of this work. The functions were implemented in Julia \cite{bezanson2017julia}.


\subsection{Variants of the GES2N objective}
\label{sec:Method:Variants}

In Figure \ref{fig:Method:OverviewOfMethod}, the cyclic order weighting matrices $\boldsymbol{C}_i, i \in \{s,n\}$ are obtained by processing baseline cyclic order matrices $\boldsymbol{C}_i^{base}$, with the processing denoted $\boldsymbol{C}_i(\boldsymbol{C}_i^{base},\boldsymbol{b})$. The following baseline cyclic order weighting matrix is used for the numerator:
\begin{equation}
	\label{eqn:Method:C_matrix:Num:Baseline}
	C_{s}^{base}[m,l] =  \left\{
	\begin{array}{ccl}
		1 & \text{ if } \left(m + 1\right) \cdot \alpha_c - \frac{1}{2} \Delta \alpha_{b} \leq \alpha[l] \leq  \left(m + 1\right) \cdot \alpha_c + \frac{1}{2} \Delta \alpha_{b} \\
		0 & \text{ otherwise } 
	\end{array}
	\right.,
\end{equation}
with $l \in \{0,1,\ldots,N_{f}-1\}$ and $m \in \{0,1,\ldots,N_{b,{s}}-1\}$. This weighting matrix is similar to the matrix in Figure \ref{fig:Method:WeightingMatrix_Sum}, i.e., it is non-zero for each cyclic order $\alpha[l]$ that is within a band of $\Delta \alpha_{b}$ around the targeted cyclic order $\alpha_c$. 

Two cyclic order weighting matrices, obtained by processing the baseline cyclic order weighting matrix, are considered for the numerator. The first cyclic order weighting matrix calculates the mean of the amplitudes in all bands and is defined as follows:
\begin{equation}
	\label{eqn:Method:C_matrix:Num:Mean}	
	{C}^{mean}_{s}[m,l] = \frac{C_{s}^{base}[m,l]}{\sum_{m=0}^{N_{b,{s}}-1} \sum_{l=0}^{N_f-1} C_{s}^{base}[m,l]},
\end{equation}
and the second cyclic order weighting matrix calculates the maximum amplitude in each targeted band: 
\begin{equation}
	\label{eqn:Method:C_matrix:Num:Max}	
	{C}_{s}^{max}[m,l] = \left\{
	\begin{array}{ccl}
		1 & \text{ if } {C}_{s}^{base}[m,l] b[l] = \underset{q}{\max} \left\{{C}_{s}^{base}[m,q] b[q]\right\} \\
		0 & \text{ otherwise } 
	\end{array}
	\right..
\end{equation}
The mean of the amplitudes in Equation \eqref{eqn:Method:C_matrix:Num:Mean}	is proportional to the amplitudes' sum and, therefore, will have the same optima. Hence, either $\boldsymbol{C}_{s}^{max}$ or $\boldsymbol{C}_{s}^{mean}$ will be used as $\boldsymbol{C}_{s}$ in Equation \eqref{eqn:Method:MainObjective}.  We found the gradient in Equation \eqref{eqn:Method:MainObjective_GradientLog_Full} performs well when using $\boldsymbol{C}_{s}^{max}$ as the numerator's cyclic order weighting matrix, despite $\boldsymbol{C}_{s}^{max}$ being a function of $\boldsymbol{b}$. 

The baseline cyclic order weighting matrix of the denominator is defined using a single band $N_{b,d} = 1$ as follows:
\begin{equation}
	\label{eqn:Method:C_matrix:Denominator:Baseline}	
	C_{n}^{base}[0,l] =  \left\{
	\begin{array}{ccl}
		1 & \text{if } \alpha_{n,min} \leq \alpha[l] \leq \alpha_{n,max} \text{ and } \sum_{m=0}^{N_{b,{s}}-1} C_{s}[m,l] = 0 \\
		0 & \text{ otherwise } 
	\end{array}
	\right.,
\end{equation}
where $\boldsymbol{C}_{n}^{base} \in \mathbb{R}^{1 \times N_f}$ and $\alpha_{n,min}$ and $ \alpha_{n,max}$ denotes the minimum and maximum cyclic order for estimating the noise.  It is non-zero for all cyclic orders in the range $\alpha_{n,min} \leq \alpha[l] \leq \alpha_{n,max}$, excluding the cyclic orders that fall in the numerator's bands, i.e., excluding cyclic orders where $\sum_{m=0}^{N_{b,{s}}-1} C_{s}[m,l] \neq 0$. More complex weighting matrices, where the noise is estimated around the targeted bands \cite{schmidt2021informative}, can be used with this formulation but are not investigated in this work. Instead, the cyclic order weighting matrix of the noise component is processed as follows:
\begin{equation}
	\label{eqn:Method:C_matrix:Den:Max}	
	{C}_{n}^{mean}[m,l] = \frac{C_{n}^{base}[m,l]}{\sum_{m=0}^{N_{b,{n}}-1} \sum_{l=0}^{N_f-1} C_{n}^{base}[m,l]},
\end{equation}
i.e., the mean of the amplitudes is calculated with $\boldsymbol{C}_{n}^{mean} = \boldsymbol{C}_{n}(\boldsymbol{C}_{n}^{base}, \boldsymbol{b})$.

In Table \ref{tab:Method:VariantsOfGES2N}, five objectives are derived using the GES2N formulation. The GES2N-ICS2 uses an objective similar to the CYCBD and ACYCBD methods. However, the cyclic order array is predefined with a fixed cyclic order resolution, and the objective uses the VS-DFT in the denominator, which provides a more consistent estimate of the mean energy of the signal under time-varying speed conditions. However, this does not improve the enhancement of the fault signatures. Since this work's focus is on fault signature enhancement and the source code of CYCBD  \cite{buzzoni2018blind} and ACYCBD  \cite{zhang2021adaptive} are available, the GES2N-ICS2 will not be investigated.

\begin{table}[h]
	\centering
	\caption{Five objective functions are derived using the proposed GES2N function. The GES2N-ICS2 uses the ICS2 as the objective. The names of the last four objectives are defined as follows: GES2N-X-Y, with $X$ indicating whether the mean of the amplitudes (Mean) or the maximum of the amplitudes (Max.) in the cyclic order bands are calculated and $Y$ indicating whether the full cyclic order range (excluding the targeted harmonics) or only a part of the cyclic order range (excluding the targeted harmonics) is used to estimate the noise or normalisation term. The latter is denoted Nf, and the former is denoted Np. The number of harmonics in the numerator of Equation \eqref{eqn:Method:MainObjective} is denoted $N_{h,{s}}$, and the target cyclic order is denoted $\alpha_c$.}
	\begin{tabular}{c c c c}  \hline
		Name & Numerator & Denominator:&  Denominator:  \\  
		& &  $\alpha_{min}$ & $\alpha_{max}$ \\  \hline
		GES2N-ICS2    & Eq. \eqref{eqn:Method:C_matrix:Num:Max} &  0.0 & 0.0 \\
		GES2N-Mean-Nf & Eq. \eqref{eqn:Method:C_matrix:Num:Mean}	 & 0.0 & $(N_{h,{s}}+1) \cdot \alpha_{c}$ \\
		GES2N-Mean-Np & Eq. \eqref{eqn:Method:C_matrix:Num:Mean}	 & 0.5 & $(N_{h,{s}}+1) \cdot \alpha_{c}$ \\
		GES2N-Max-Nf  & Eq. \eqref{eqn:Method:C_matrix:Num:Max}	 & 0.0 & $(N_{h,{s}}+1) \cdot \alpha_{c}$ \\
		GES2N-Max-Np  & Eq. \eqref{eqn:Method:C_matrix:Num:Max}	 & 0.5 & $(N_{h,{s}}+1) \cdot \alpha_{c}$ \\	 \hline			
	\end{tabular}
	\label{tab:Method:VariantsOfGES2N}
\end{table}

The GES2N-Mean-Nf objective is consistent with the MSESHIRD method \cite{zhou2022blind}. Still, instead of summing all the amplitudes in the bands in the numerator, the mean of the amplitudes is calculated. Furthermore, only the cyclic orders associated with the first $10$ harmonics are utilised in the denominator, not the full envelope spectrum. The $0$ cyclic order component calculates the squared energy of the angle domain signal and can be sensitive to extraneous components. Hence, the GES2N-Mean-Np objective is also investigated, which ignores the low cyclic order components (below 0.5 cyclic orders). The mean in the numerator can be insensitive to sparse components. Therefore, the maximum amplitudes in each cyclic order band are utilised to extract the amplitudes in the GES2N-Mean-Nf and the GES2N-Mean-Np objectives. 

The parameters for the proposed GES2N variants used are as follows unless stated otherwise: $\Delta \alpha_{b} = 0.1$ shaft orders, $N_{h,{s}} = 10$, and $\Delta \alpha$ is equal to the cyclic order resolution of the order tracked signal's SES. Borghesani et al. \cite{borghesani2014velocity} provide further details about the required cyclic order resolution for the VS-DFT.

\subsection{Summary of analyses}
\label{sec:Method:Summary}

In this section, the optimal filter design methods considered, the optimisation parameters used, and the quantitative metrics of the SES used to quantify the methods' performance are summarised.

\subsubsection{Comparison with other methods}
\label{sec:Method:Summary:ComparisonWithOtherMethods}
The GES2N-Mean-Nf, GES2N-Mean-Np, GES2N-Max-Nf, and GES2N-Max-Np in Table \ref{tab:Method:VariantsOfGES2N} are compared to the CYCBD \cite{buzzoni2018blind}, ACYCBD \cite{zhang2021adaptive}, MOMEDA \cite{mcdonald2017multipoint}, the $L_2$/$L_1$ norm of the envelope spectrum \cite{peeters2020blind} and the spectral negentropy of the SES \cite{peeters2020blind}. For the CYCBD, the angle domain formulation was utilised with the first $10$ harmonics used and a cyclostationary order of $2$ was maximised. For the ACYCBD, the angle formulation version was utilised due to the time-varying speed conditions. A cyclostationary order of $2$ was maximised with no frequency limit imposed. MOMEDA was applied to the time domain signal, the average gear rotational speed was used as the cyclic frequency, and the default windowing function was utilised. 

\subsubsection{Optimisation parameters}
\label{sec:Method:Summary:OptimisationParameters}
The following parameters are used for all the algorithms unless stated otherwise: Convergence tolerance: $10^{-12}$, Maximum number of iterations: $1500$, Filter initialisation method: Linear predictive coding \cite{buzzoni2018blind} (except for MOMEDA, which is a non-iterative method). The filter length $D = 256$ for all algorithms unless stated otherwise. 

\subsubsection{Quantitative metrics of the SES}
\label{sec:Method:Summary:QuantitativeMetrics}
The SES of the filtered signals and the filter coefficients' frequency response functions are compared between the different methods. To quantitatively evaluate and compare the performance of the methods, the following four quantitative performance metrics of the SES of the filtered signals are used in subsequent sections:
\begin{itemize}
    \item[M1:] The first metric calculates the ratio between the mean of the amplitudes of the first $N_{h}$ targeted harmonics in the SES and the median of the SES. Hence, it is a different estimate of the signal-to-noise ratio in the SES and is large if the damage is prominent relative to the noise floor in the SES.

    \item [M2:] The second metric calculates the ratio between the mean of the amplitudes of the first $N_{h}$ targeted harmonics in the SES and a known dominant extraneous component in the signal. Hence, this metric is large if the targeted amplitudes are more prominent than the extraneous component. 

    \item [M3:] The third metric calculates the ratio between the mean of the amplitudes of the first $N_{h}$ targeted harmonics in the SES and the maximum amplitude in the SES in the cyclic order range between $0.5$ and $20 \cdot \alpha_c$. Hence, this metric is close to $1$ if the metric only contains information related to the targeted component and there are small variations in the amplitudes of the components. 

    \item [M4:] The last metric calculates the precision (i.e., reciprocal of the variance) of the amplitudes of the first $N_{h}$ targeted harmonics in the SES. This metric estimates the variation in the component's amplitudes in the SES.
\end{itemize}

\section{Experimental results}
\label{sec:ExpData:main}

The methods are compared on three experimental datasets. In Section \ref{sec:ExpData:Setup}, an overview of the experimental test bench is presented, whereafter the results of the three datasets are presented in Sections \ref{sec:ExpData:Dataset:LGD} - \ref{sec:ExpData:Dataset:DGD2}. The impact of the hyperparameters on the proposed objective function is briefly demonstrated in Section \ref{sec:ExpData:Dataset:SensitivityAnalysis}, whereafter the results are discussed in  Section \ref{sec:ExpData:Dataset:Discussion}.

\subsection{Overview of setup}
\label{sec:ExpData:Setup}
\begin{figure}[h]
	\centering
	\includegraphics[width=0.95\linewidth]{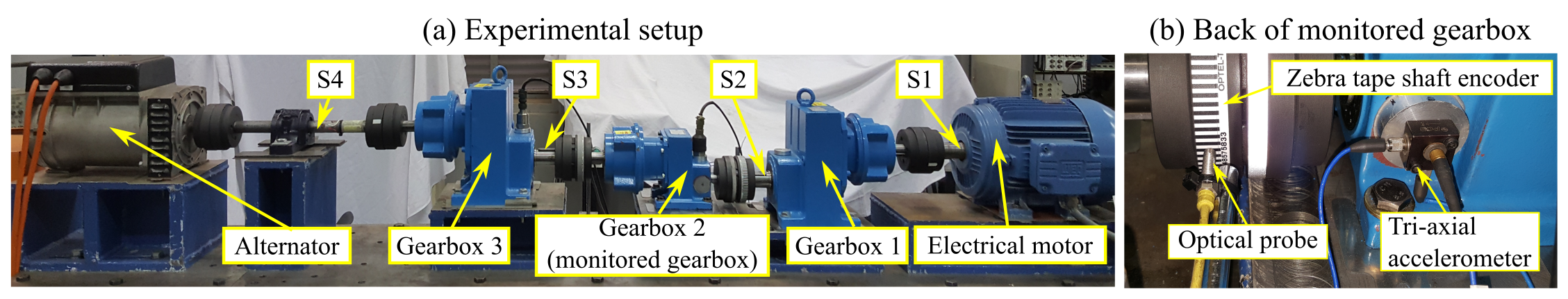}
	\caption{The helical gearbox test bench in the Centre for Asset Integrity Management (C-AIM) laboratory is presented. (a) The components in the test bench are shown. (b) The sensors that were utilised are shown. }
	\label{fig:ExpData:Setup}
\end{figure}

The helical gearbox test bench in the Centre for Asset Integrity Management (C-AIM) laboratory of the University of Pretoria is used. The test bench, shown in Figure \ref{fig:ExpData:Setup}, consists of a 5.5 kW electrical motor that drives an alternator. Three helical gearboxes are used in the system, with the centre gearbox (gearbox 2) instrumented with accelerometers. The axial component of the tri-axial accelerometer, shown in Figure \ref{fig:ExpData:Setup}(b), is used in subsequent sections. An optical probe is used on an 88 pulses per revolution zebra tape shaft encoder on gearbox 2's input shaft, as shown in Figure \ref{fig:ExpData:Setup}. The gear connected to the gearbox 2's input shaft is damaged, so modulation from the gear damage is expected at $1.0$ shaft orders. Hence, $\alpha_c = 1.0$ for this dataset.
The optical probe signal is sampled at 51.2 kHz, and the tri-axial accelerometer signal is sampled at 25.6 kHz using an OROS OR35 data acquisition system. This work uses the tri-axial accelerometer's axial component. 

The following three datasets are considered: Localised gear damage, distributed gear damage (case 1), and distributed gear damage (case 2). For the localised gear damage case in Section \ref{sec:ExpData:Dataset:LGD}, a slot is seeded in a tooth of the gear. For the distributed gear damage cases, gear damage is distributed across the faces of multiple gear teeth, with case 2's damage in Section \ref{sec:ExpData:Dataset:DGD2} being more severe than case 1's damage in Section \ref{sec:ExpData:Dataset:DGD1}.

In the next section, the dataset with localised gear damage is considered. 

\subsection{Localised gear damage}
\label{sec:ExpData:Dataset:LGD}
\begin{figure}[h]
	\centering
	\begin{subfigure}{0.37\linewidth}
		\caption{}
		\includegraphics[width=0.96\linewidth]{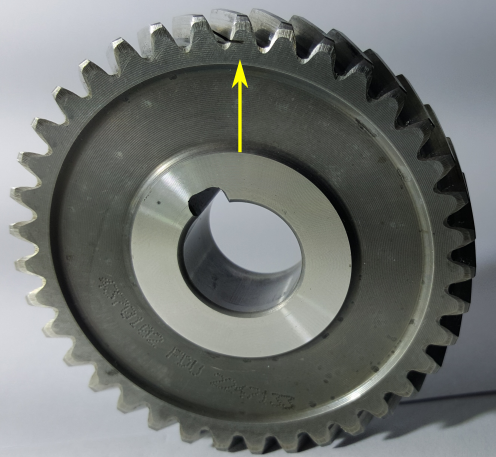}
		\label{fig:Results:Exp:LGD:Gear:Before}		
	\end{subfigure}
	\begin{subfigure}{0.37\linewidth}
		\caption{}
		\includegraphics[width=0.96\linewidth]{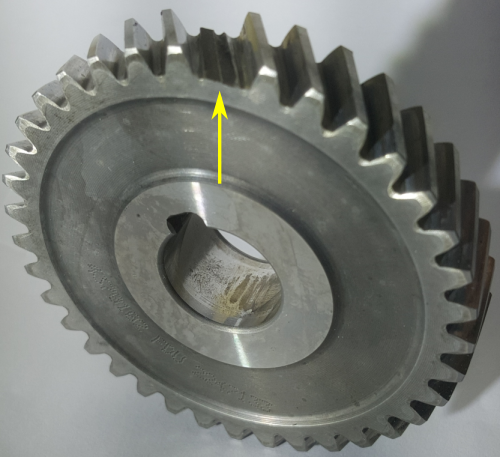}
		\label{fig:Results:Exp:LGD:Gear:After}				
	\end{subfigure}
	\caption{The gear with localised damage is shown in \ref{fig:Results:Exp:LGD:Gear:Before} before the experiment was started and in \ref{fig:Results:Exp:LGD:Gear:After} after the experiment was completed.}
	\label{fig:Results:Exp:LGD:Gear}
\end{figure}

\begin{figure}[h]
	\centering
	\begin{subfigure}{0.287\linewidth}
		\centering 
		\caption{Meas. 1}
		\includegraphics[width=0.99\linewidth]{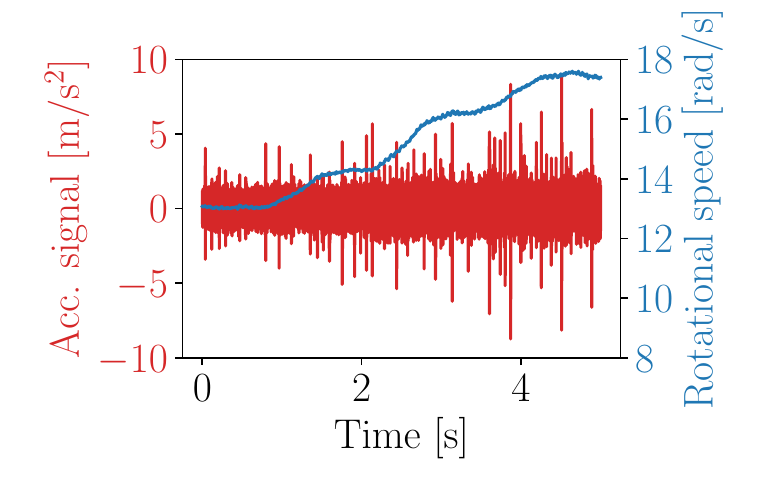}
		\label{fig:Results:Exp:LGD:Measurements:Meas1}		
	\end{subfigure}
	\begin{subfigure}{0.287\linewidth}
		\centering 
		\caption{Meas. 2}
		\includegraphics[width=0.99\linewidth]{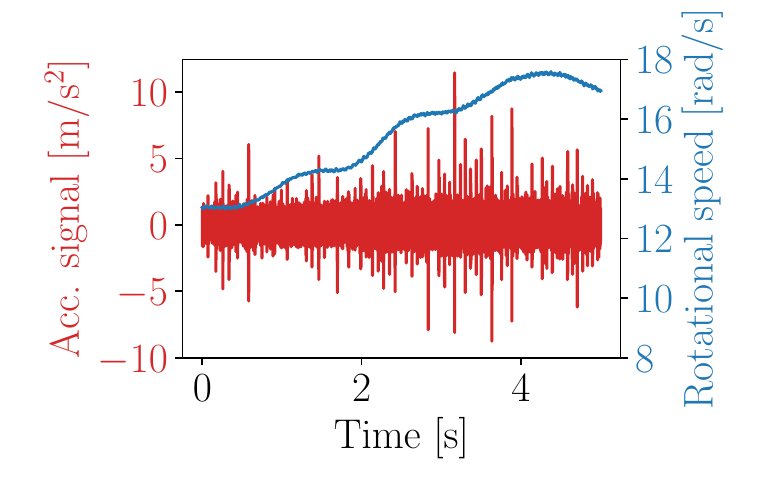}
		\label{fig:Results:Exp:LGD:Measurements:Meas2}		
	\end{subfigure}	
	\begin{subfigure}{0.287\linewidth}
		\centering 
		\caption{Meas. 3}
		\includegraphics[width=0.99\linewidth]{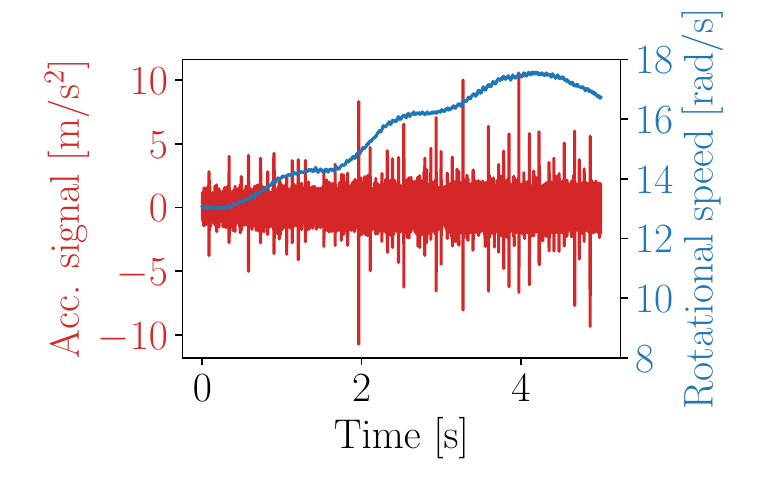}
		\label{fig:Results:Exp:LGD:Measurements:Meas3}		
	\end{subfigure}	
	\begin{subfigure}{0.287\linewidth}
		\centering 
		\caption{Meas. 4}
		\includegraphics[width=0.99\linewidth]{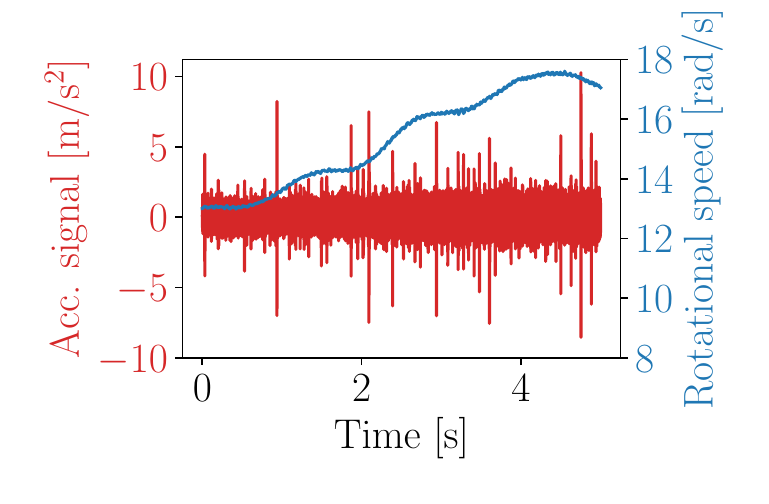}
		\label{fig:Results:Exp:LGD:Measurements:Meas4}		
	\end{subfigure}	
	\begin{subfigure}{0.287\linewidth}
		\centering 
		\caption{Meas. 5}
		\includegraphics[width=0.99\linewidth]{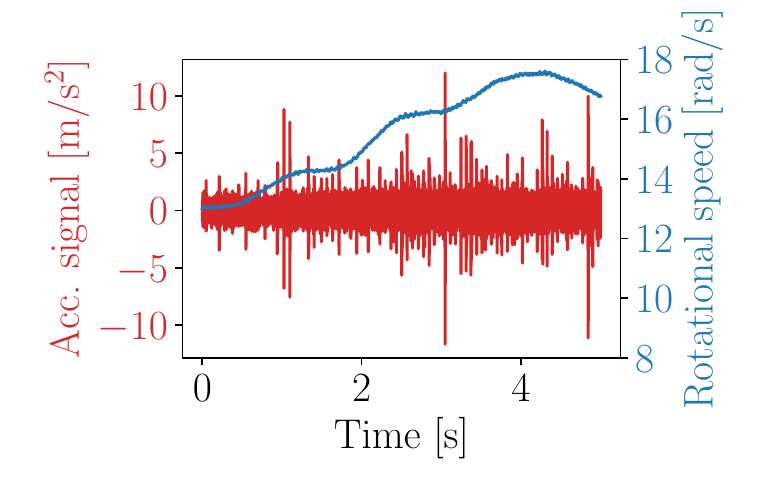}
		\label{fig:Results:Exp:LGD:Measurements:Meas5}		
	\end{subfigure}	
	\caption{The five measurements acquired from the gearbox test bench are shown for the localised gear damage case. The time domain vibration signal is shown in red with a corresponding $y$-axis on the left side, while the rotational speed signal is in blue and has a corresponding $y$-axis on the right side.}
	\label{fig:Results:Exp:LGD:Measurements}
\end{figure}

The experimental test rig acquired data from a gear with localised gear damage. 
The slot, shown in Figure \ref{fig:Results:Exp:LGD:Gear:Before}, was seeded in one of the teeth of gearbox 2's gear before the tests were started. After the tests were completed, the gear is shown in Figure \ref{fig:Results:Exp:LGD:Gear:After}. Five signals with a duration of $5$ seconds, acquired over the life of the gear, are considered from this dataset. The gearbox was not disassembled or inspected during the experimental study. Therefore, the exact condition of the gearbox for the measurements is unknown, but the severity of the damage was the least for measurement $1$ and the most for measurement $5$. Figure \ref{fig:Results:Exp:LGD:Measurements} shows these five acceleration and rotational speed signals. The system operates under quasi-stationary regimes, with an approximate linear ramp-up between the regimes as shown in Figure \ref{fig:Results:Exp:LGD:Measurements}.  Dominant extraneous impulses in the time domain signal manifest in the raw signals' SES. These extraneous impulses occurred at approximately $5.72$ shaft orders and were in the healthy and damaged gearbox signals. These components impede the performance of conventional time domain and envelope spectrum-based metrics. 

\begin{figure}[h]
	\centering
	\includegraphics[width=0.98\linewidth]{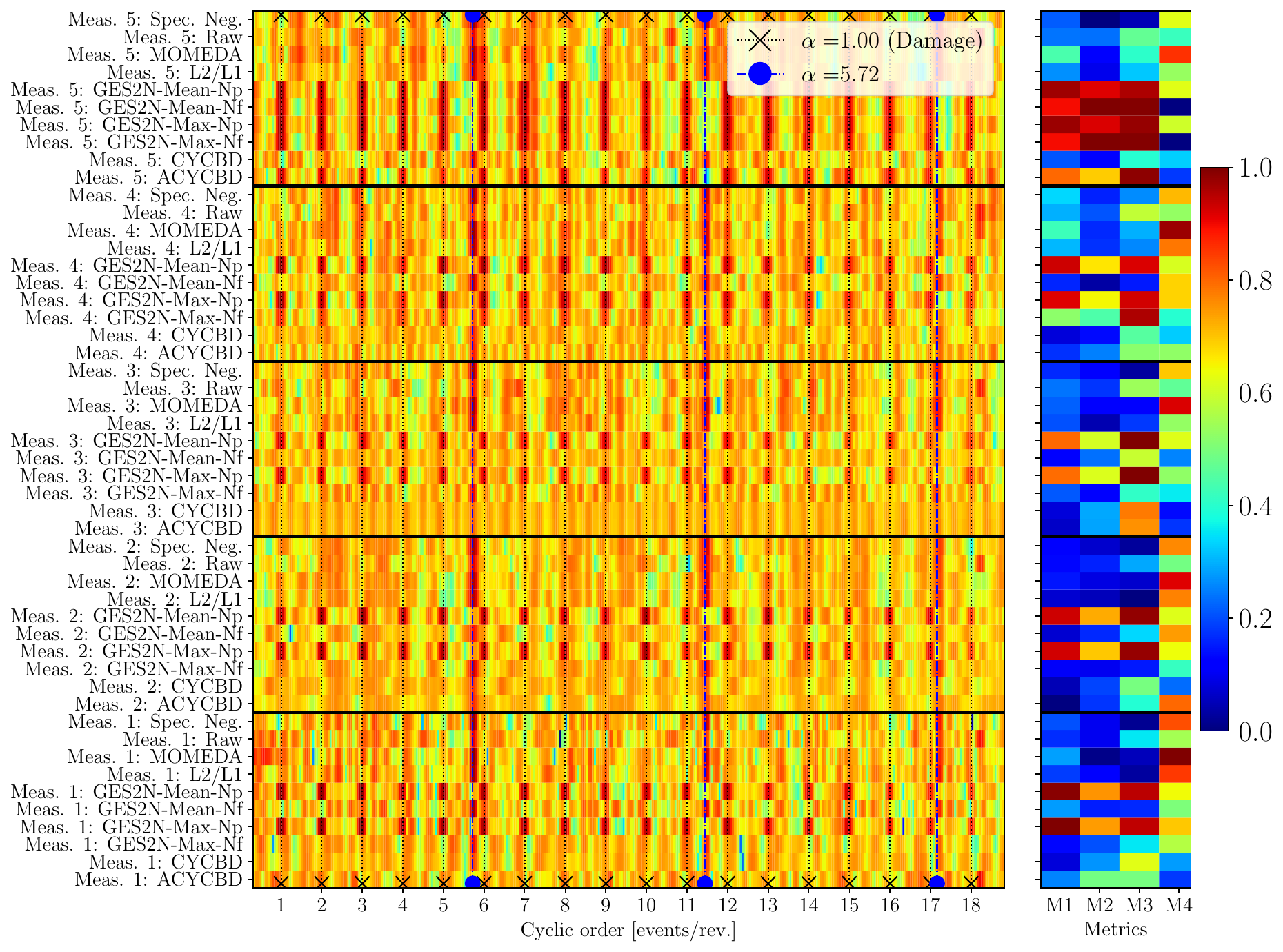}
	\caption{The logarithm of the median normalised Squared Envelope Spectra (SESa) of the filtered signals $\boldsymbol{y}$ for the gear in Figure \ref{fig:Results:Exp:LGD:Gear} are presented on the left side with the logarithm of the normalised SESa of the raw signals (denoted Raw). The SESa are normalised with their respective medians. The damage manifests at $\alpha = 1.0$ and its harmonics, and $5.72$ shaft orders and its harmonics are associated with extraneous impulsive components.  The four performance metrics, described in Section \ref{sec:Method:Summary:QuantitativeMetrics}, are shown on the right. Each respective metric is normalised between $0$ and $1$ across the rows. The SES plot on the left side is also normalised to have a magnitude between $0$ and $1$, and therefore, both plots use the same colour bar.}
	\label{fig:Results:Exp:LGD:SES_X}
\end{figure}

The objective functions were optimised as described in Section \ref{sec:Method:main} for the five signals to obtain optimal filter coefficients. After that, the filtered signals were obtained, and the Squared Envelope Spectra (SESa) were calculated. The SESa of the filtered and raw signals are shown in Figure \ref{fig:Results:Exp:LGD:SES_X}. To condense the results, the SESa are presented as three-dimensional plots in Figure \ref{fig:Results:Exp:LGD:SES_X} for the five signals; the $x$-axis shows the cyclic orders, the $y$-axis shows the measurement number and the name of the objective, and the colours quantify the logarithm of the median normalised SES's amplitude. The SESa were normalised to make them easier to compare. In Figure \ref{fig:Results:Exp:LGD:SES_X}, two cyclic order components are highlighted, namely, the gear's fundamental component ($\alpha = 1.0$) and its harmonics and the extraneous impulsive component  ($\alpha = 5.72$) and its harmonics. The extraneous component was also present in the healthy measurements, and the exact source is unknown. However, we speculate that the floating bearing in the monitored gearbox causes this.

Four metrics, discussed in Section \ref{sec:Method:Summary:QuantitativeMetrics}, are also summarised in Figure \ref{fig:Results:Exp:LGD:SES_X} to quantify the properties of the SES. The metrics were normalised within the range of $0$ to $1$ along the rows, while the logarithm of the normalised SESa was also adjusted to fall within the same range of $0$ to $1$, ensuring uniformity in colour representation.

\begin{figure}[h]
	\centering
	\includegraphics[width=0.98\linewidth]{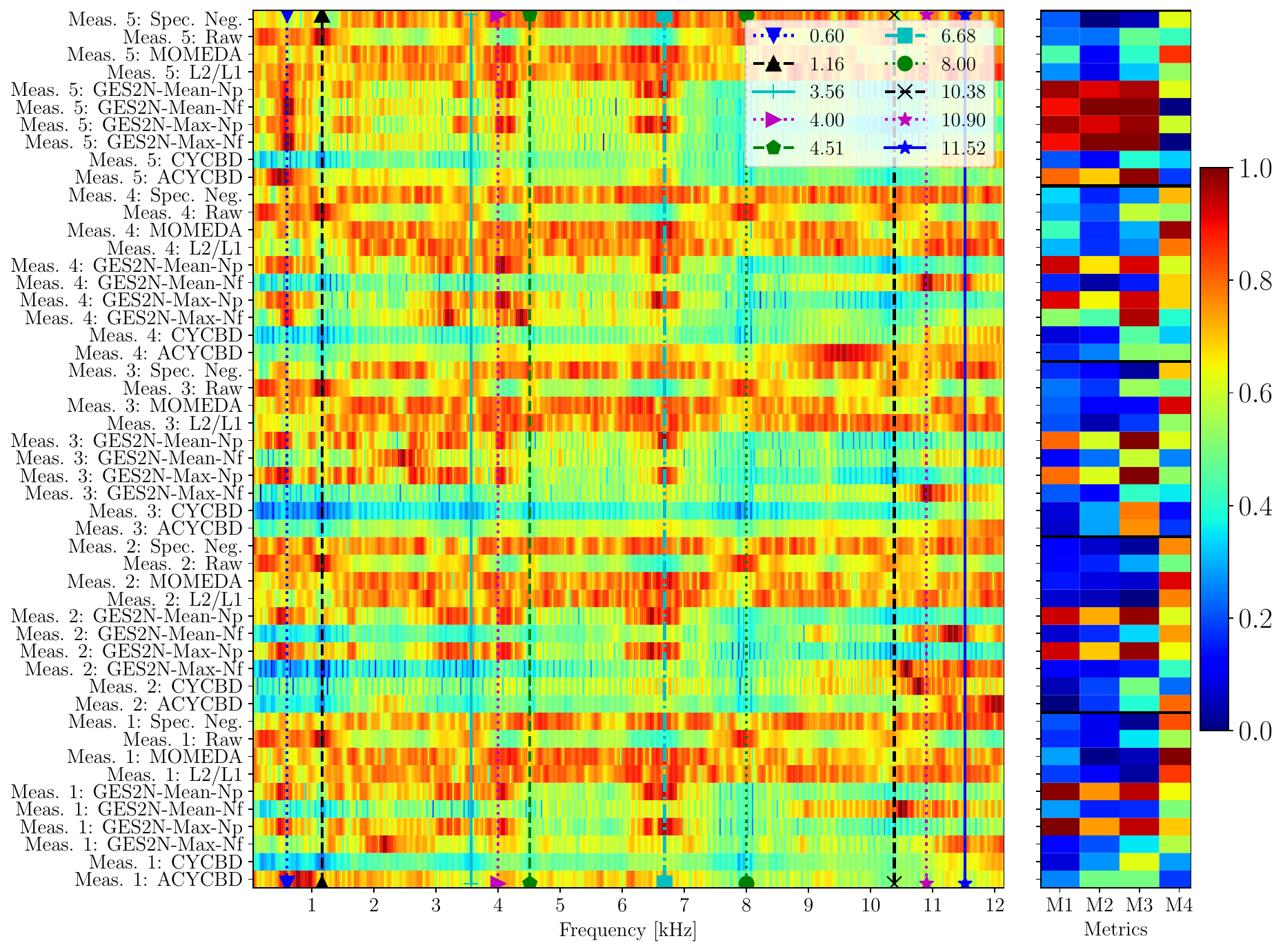}
	\caption{The normalised spectra of the optimised filter coefficients are presented. The figures refer to the normalised power spectral densities of the raw signals as \textit{Raw}. The $L_2$-norm of the spectrum was used to normalise the respective spectrum. Dominant frequencies (in kHz) are identified and marked in the legend to aid the discussion. The metrics are repeated here for the reader's convenience. }
	\label{fig:Results:Exp:LGD:FRF_H}
\end{figure}

The magnitude of the $L_2$-normalised optimal filter coefficients' spectra can be used to understand which spectral frequency bands' content maximises the objective function (i.e., which bands contain the diagnostic information). The magnitude of the optimal filter coefficients' spectra is shown in Figure \ref{fig:Results:Exp:LGD:FRF_H} with the power spectral densities of the raw signals also shown as a baseline. Some dominant frequency bands are highlighted with vertical markers.

The raw signals are dominated by the extraneous component at $5.72$ orders, and the gear damage components at $1.0$ order and its harmonics are hidden as seen in the SESa and metrics' results in Figure \ref{fig:Results:Exp:LGD:SES_X}. In contrast, the gear damage symptoms are visible in the SESa of all measurements when using the filter coefficients obtained with the GES2N-Max-Np and the GES2N-Mean-Np-based objective functions. Both objectives' filtered results contain dominant components at $1.0$ orders and their harmonics and large scores for the M1, M2 and M3 metrics. The GES2N-Max-Nf-based objective function can only enhance the gear damage components for measurements $4$ and $5$ (i.e. when the damage is more pronounced). Furthermore, the GES2N-Mean-Nf objective function only enhanced the gear damage in measurement $5$ (i.e., for the most severe damage case). 

When comparing the metrics obtained using these objective functions, the GES2N-Max-Np and GES2N-Max-Np are better than GES2N-Max-Nf and GES2N-Max-Nf for early fault detection. Also, they have better signal-to-noise ratios (M1) for the last measurement and generally lower variances (or a higher precision) in the component's amplitudes in the SES (M4). However, the GES2N-Max-Nf and GES2N-Max-Nf can perform slightly better to enhance the damage component relative to the extraneous component (i.e., M2), and the damage components are the most prominent in the SES (M3) for the last measurement. Furthermore, measurement 4 highlights that the GES2N-Max-Nf performs slightly better than the GES2N-Mean-Nf, with M1 - M3 being higher and M3 being the largest of all objectives.

The raw signals' power spectral densities in Figure \ref{fig:Results:Exp:LGD:FRF_H} (denoted Raw) highlight multiple high-energy frequency bands.
The spectra of the best-performing filter coefficients (i.e., GES2N-Max-Np and the GES2N-Mean-Np) in Figure \ref{fig:Results:Exp:LGD:FRF_H} contain multiple dominant frequency bands that are not dominant in the power spectral density of the raw signal (denoted Raw). Hence, the damage does not manifest in the dominant frequency bands. Furthermore, all the bands enhanced by the GES2N-Max-Np and GES2N-Max-Np objective functions contain fault information, but the fault information is most prominent in the $0.6$ kHz band.

From the results in Figure \ref{fig:Results:Exp:LGD:SES_X}, the CYCBD objective function did not perform well for any of the measurements, whereas the ACYCBD was able to enhance the gear damage components for measurement $5$ (i.e., the most severe damage case). Compared to the best-performing filters, the CYCBD and ACYCBD tend to focus on the higher spectral frequencies in Figure \ref{fig:Results:Exp:LGD:FRF_H}. The extraneous $5.72$ order component dominates these higher frequency bands. The GES2N-Max-Nf, the GES2N-Mean-Nf, CYCBD and the ACYCBD, which all contain the mean squared energy of the filtered signals in their denominator, typically enhance spectral frequency content above $10$ kHz, which are dominated by extraneous impulsive behaviour. This indicates that the performance of the objective functions is impeded by including the lower cyclic order components (i.e. mean squared energy of the filtered signal) in the denominator of the objective function.

It is not possible to detect the damaged components in Figure \ref{fig:Results:Exp:LGD:SES_X} using MOMEDA. The SESa corresponding to MOMEDA contains the smallest variation in the targeted component's amplitude for all the cases, because the damage is not prominent in the signal. The MOMEDA filter coefficients' spectra in Figure \ref{fig:Results:Exp:LGD:FRF_H} show that the filter focused on similar spectral frequency bands as the GES2N-Max-Np and the GES2N-Mean-Np functions, except that MOMEDA does not enhance the component at around $0.6$ kHz. This indicates that the cyclic order at around $0.6$ kHz contains important diagnostic information. This is corroborated by the GES2N-Max-Nf filter coefficients' spectrum for measurement $5$ in Figure \ref{fig:Results:Exp:LGD:FRF_H}, which shows that only a single spectral frequency band around $0.6$ kHz is utilised to enhance the gear damage components in Figure \ref{fig:Results:Exp:LGD:SES_X}.

Lastly, blind objectives are also used in this study: The $L_2$/$L_1$ of the envelope spectrum and spectral negentropy of the SES objective functions were unable to enhance the gear damage components but enhanced the cyclic order component at $5.72$ and its harmonics. This makes sense as the component is a dominant sparse component in the SES; therefore, these methods achieve their objective. The corresponding filter coefficients' frequency response functions in Figure \ref{fig:Results:Exp:LGD:FRF_H} show that multiple frequency bands are enhanced throughout the entire frequency range. In some cases (e.g., $L_2$/$L_1$ measurement $5$, spectral negentropy measurement $3$) contains some evidence that the amplitudes in the band around $0.6$ kHz are amplified. Still, due to the other amplified bands, the damage cannot be detected.

The next section considers the first distributed gear damage case. 

\subsection{Distributed gear damage: Case 1}
\label{sec:ExpData:Dataset:DGD1}

\begin{figure}[h]
	\centering
	\begin{subfigure}{0.2\linewidth}
		\caption{}
		\includegraphics[width=0.96\linewidth]{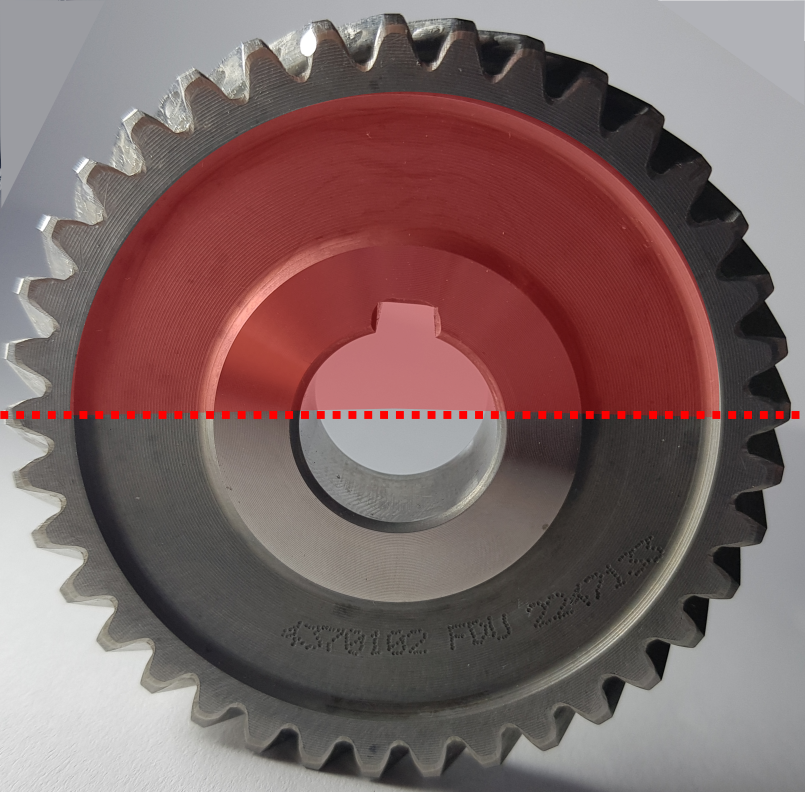}
		\label{fig:Results:Exp:DGD1:Gear:Region}		
	\end{subfigure}
	\begin{subfigure}{0.79\linewidth}
		\caption{}
		\includegraphics[width=0.96\linewidth]{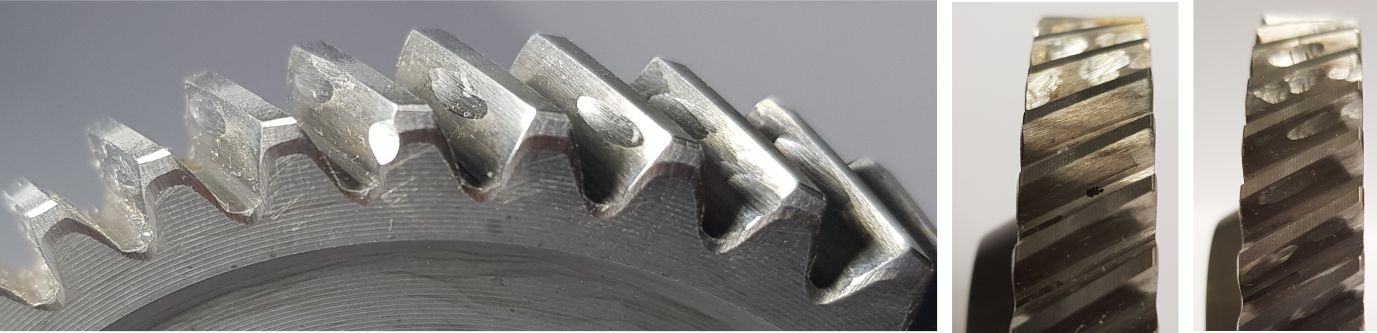}
		\label{fig:Results:Exp:DGD1:Gear:Damage}		
	\end{subfigure}
	\caption{The gear with distributed damage (case 1) is shown. (a) The section of teeth that are damaged is highlighted in red, i.e., half the gear's teeth are damaged. (b) The damaged teeth are shown.  }
	\label{fig:Results:Exp:DGD1:Gear}
\end{figure}

\begin{figure}[h]
	\centering
	\begin{subfigure}{0.35\linewidth}
		\centering 
		\caption{Meas. 1}
		\includegraphics[width=0.99\linewidth]{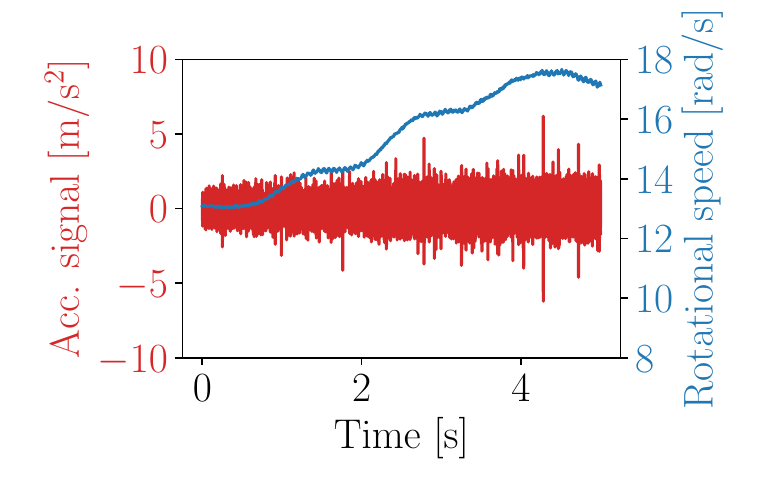}
		\label{fig:Results:Exp:DGD1:Measurements:Meas1}		
	\end{subfigure}
	\begin{subfigure}{0.35\linewidth}
		\centering 
		\caption{Meas. 2}
		\includegraphics[width=0.99\linewidth]{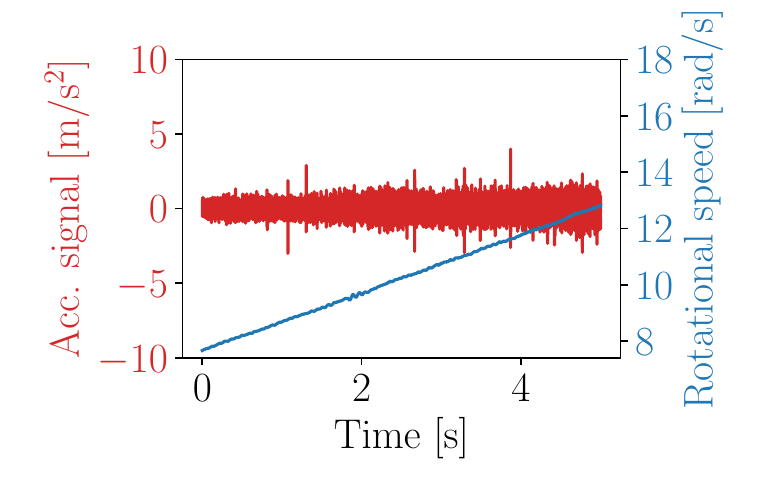}
	\label{fig:Results:Exp:DGD1:Measurements:Meas2}			
	\end{subfigure}	\\
	\begin{subfigure}{0.35\linewidth}
		\centering 
		\caption{Meas. 3}
		\includegraphics[width=0.99\linewidth]{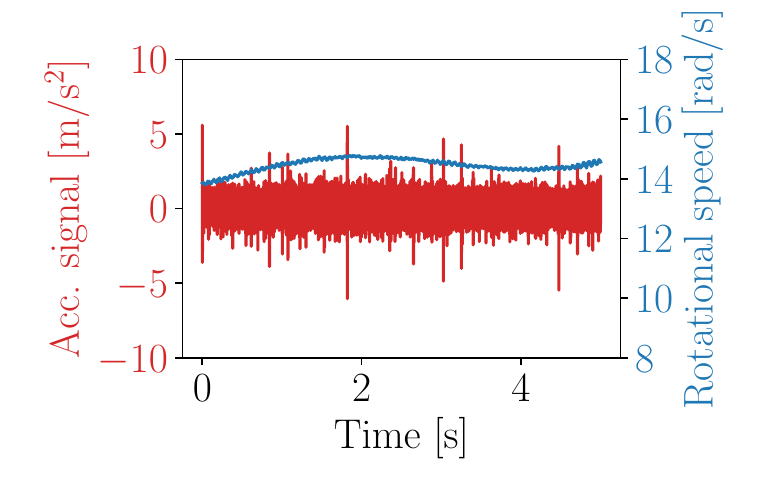}
		\label{fig:Results:Exp:DGD1:Measurements:Meas3}		
	\end{subfigure}	
	\begin{subfigure}{0.35\linewidth}
		\centering 
		\caption{Meas. 4}
		\includegraphics[width=0.99\linewidth]{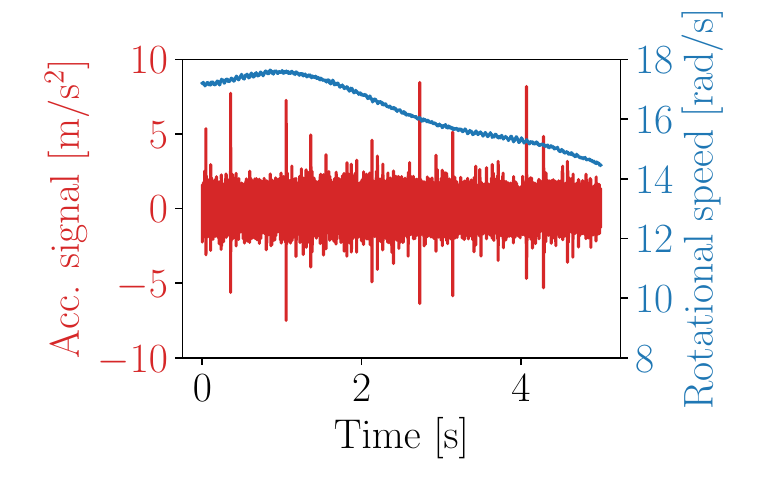}
		\label{fig:Results:Exp:DGD1:Measurements:Meas4}				
	\end{subfigure}	
	\caption{Four measurements acquired from the gearbox with the gear's condition shown in Figure \ref{fig:Results:Exp:DGD1:Gear}. The time domain vibration signal is shown in red, with its $y$-axis shown on the left side and the corresponding rotational speed of the input shaft is shown in blue with a corresponding $y$-axis on the right side. }
	\label{fig:Results:Exp:DGD1:Measurements}
\end{figure}

Experimental measurements were acquired from the gear shown in Figure \ref{fig:Results:Exp:DGD1:Gear}. The damage was distributed over half of the gear teeth as shown by the highlighted section in Figure \ref{fig:Results:Exp:DGD1:Gear:Region}. The damage was introduced across the respective teeth's faces with a Dremel rotary tool, and the teeth did not deteriorate for the considered measurements. The damaged teeth are shown in Figure \ref{fig:Results:Exp:DGD1:Gear:Damage}. Figure \ref{fig:Results:Exp:DGD1:Measurements} shows the four measurements considered in this dataset, each with its own distinct rotational speed profile. The first measurement's speed profile is similar to the speed profiles in the previous section; the second measurement contains a linear ramp-up, the third measurement contains a slight speed fluctuation, and the last one considers a predominantly speed-down scenario. Since the gear teeth are in the same condition, the results will provide insights into the effect of speed on the performance of the methods.  The impulses with a cyclic order of $5.72$ are visible in the raw signal and more dominant than the weak damage components and, therefore, are expected to impede the performance of the designed filters. 

\begin{figure}[h]
	\centering
	\includegraphics[width=0.98\linewidth]{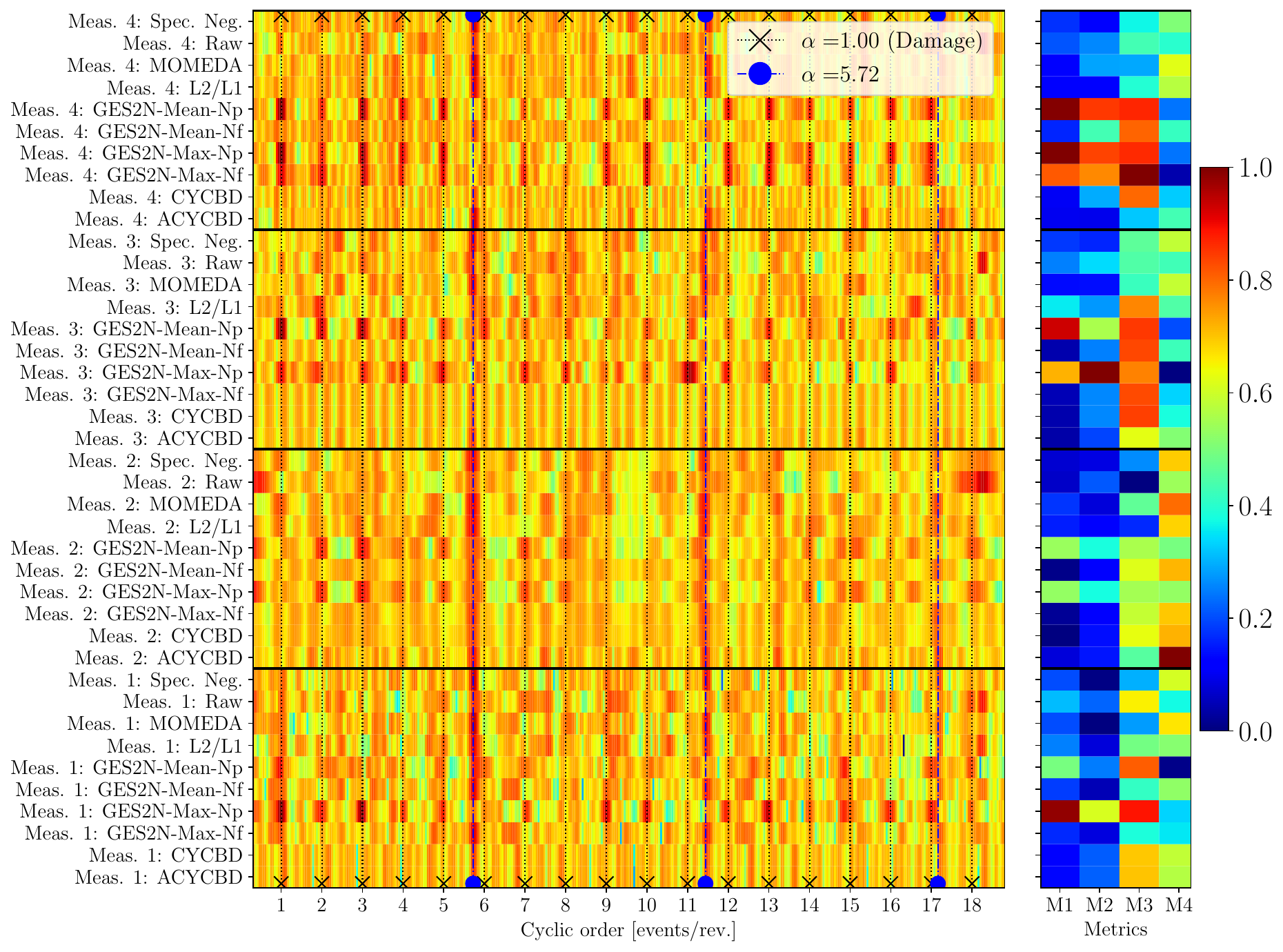}
	\caption{The logarithm of the normalised Squared Envelope Spectra (SESa) of the filtered signals $\boldsymbol{y}$ for the gear in Figure \ref{fig:Results:Exp:DGD1:Gear} are presented with the logarithm of the normalised SESa of the raw signals (denoted Raw). The same representation is used in Figure \ref{fig:Results:Exp:LGD:SES_X}. }
	\label{fig:Results:Exp:DGD1:SES_X}
\end{figure}

The previous section applied the same procedure to obtain the SESa results. The logarithm of the median normalised SESa of the filtered signals is shown in Figure \ref{fig:Results:Exp:DGD1:SES_X} for the different measurements with the logarithm of the SESa of the raw signals also shown as a baseline. The performance metrics described in Section \ref{sec:Method:Summary:QuantitativeMetrics} are also shown on the right of Figure \ref{fig:Results:Exp:DGD1:SES_X}. The SES of the raw signal contains weak harmonics related to the gear components for measurements $3$ and $4$. However, observing the damage in measurements $1$ and $2$ is challenging. 

From a fault detection perspective, the GES2N-Max-Np performed the best out of all the objective functions on this dataset since the gear damage is enhanced in all the measurements. The GES2N-Mean-Np performed only slightly worse as it could only enhance the first harmonic of the damaged component in measurement $1$, which is also corroborated by the M1, M2 and M3 metrics. However, according to the M1 metric, the signal-to-noise of the damaged components is more enhanced through the GES2N-Mean-Np objective function for measurement $3$. Even though the components are more prominent, the GES2N-Max-Np can enhance the damage better relative to the extraneous component (M2). Measurement $2$ is the most challenging, with only some gear damage harmonics visible when using the GES2N-Max-Np and GES2N-Mean-Np filtered signals. The metrics corroborate this; metrics M1, M2, and M3 show some evidence of damage, but it is less convincing than the other measurements. The GES2N-Max-Nf could only enhance the damaged gear components in measurement $4$, while the GES2N-Mean-Nf could not enhance the fault signatures in any of the considered measurements. This further highlights that using the maximum instead of the mean can be better for weak fault detection.

The CYCBD, ACYCBD, MOMEDA, the spectral negentropy and the $L_2$/$L_1$ objective functions could not enhance the damaged components in the signal, with only the dominant component at $5.72$ shaft orders and its harmonics being visible in the SESa plots. For brevity's sake, the frequency spectra of the filter coefficients are not included. However, the results corroborate the previous dataset's results; the best-performing filter coefficient's spectra contain dominant spectral frequencies that do not align with the power spectral density dominant frequency bands, highlighting that the gear damage manifests in weaker spectral frequency bands. 

\subsection{Distributed gear damage: Case 2}
\label{sec:ExpData:Dataset:DGD2}
\begin{figure}[h]
	\centering
	\includegraphics[width=0.7\linewidth]{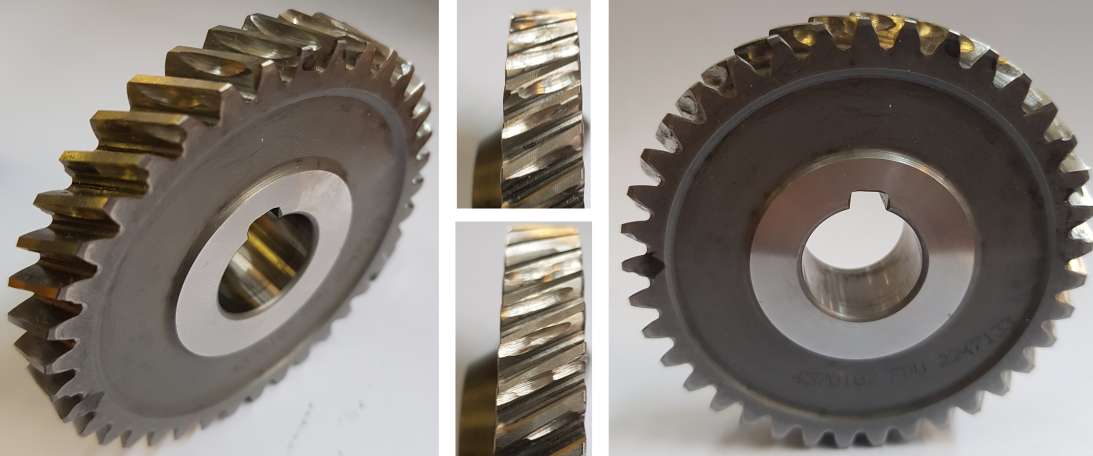}
	\caption{The gear with distributed damage (case 2) is shown. The gear in Figure \ref{fig:Results:Exp:DGD2:Gear} was reused, but the teeth's damage increased.}
	\label{fig:Results:Exp:DGD2:Gear}
\end{figure}

The second distributed gear damage dataset was obtained with the gear shown in Figure \ref{fig:Results:Exp:DGD2:Gear}. This gear is the same gear considered in the previous section, but the severity of the damaged teeth was further increased with a Dremel rotary tool, i.e., the healthy teeth were not damaged. Hence, the gear damage is much more prominent than the previous dataset. Four measurements are considered in this section with the same rotational speed profiles as the previous dataset, and therefore, the time domain signals are not included for brevity's sake.  Since the gear damage is much more prominent than the distributed gear damage in case 1, the different objective functions are expected to perform much better on this dataset.

\begin{figure}[h]
	\centering
	\includegraphics[width=0.98\linewidth]{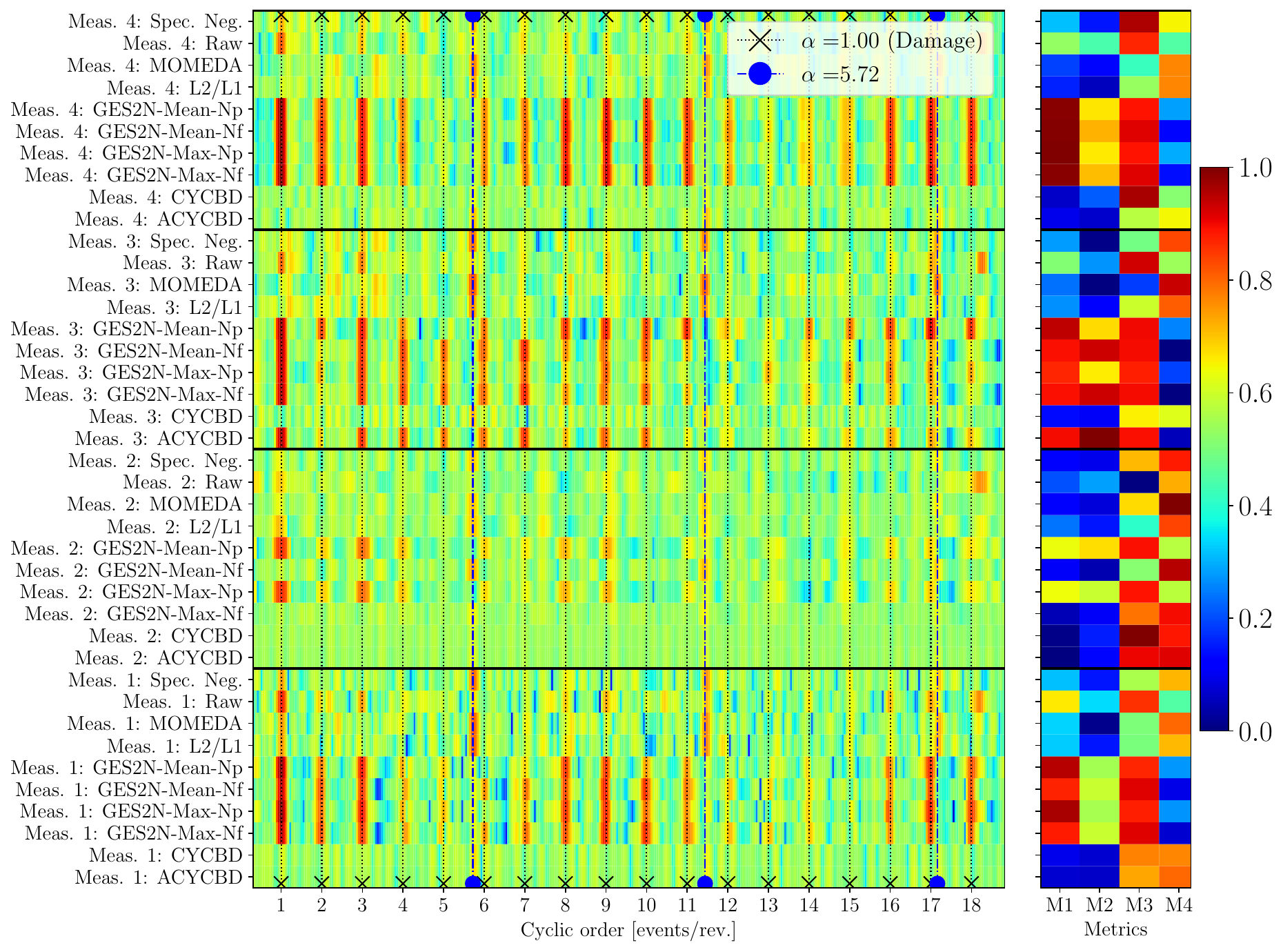}
	\caption{The logarithm of the normalised Squared Envelope Spectra (SESa) of the filtered signals $\boldsymbol{y}$ for the gear in Figure \ref{fig:Results:Exp:DGD2:Gear} are presented with the logarithm of the normalised SESa of the raw signals (denoted Raw). The damage manifests at $\alpha = 1.0$ and its harmonics. The same representation is used as Figure \ref{fig:Results:Exp:LGD:SES_X}. }
	\label{fig:Results:Exp:DGD2:SES_X}
\end{figure}

The logarithm of the normalised SESa of the filtered signals is compared to the logarithm of the normalised SESa of the raw signals for the four measurements in Figure \ref{fig:Results:Exp:DGD2:SES_X}. The raw signal shows some damage symptoms in measurement $1$, measurement $3$ and measurement $4$, with the $1.0$ shaft order and some of its harmonics being visible. 

The GES2N-Max-Np and the GES2N-Mean-Np perform the best, with the damaged components' amplitudes prominent in the SESa for all measurements. The metrics corroborate this; the GES2N-Max-Np and GES2N-Mean-Np have similar metrics for M1, M2 and M3 for all the measurements, except for measurement $3$. In measurement $3$, the GES2N-Mean-Np enhances the signal-to-noise ratio of the damage slightly better. The GES2N-Max-Nf and the GES2N-Mean-Nf detected the damage in measurements $1$, $3$ and $4$ and performed slightly better in enhancing the damage components' amplitude relative to the extraneous components (M2). However, the extraneous component is enhanced in measurement $2$, and the methods fail to detect damage. Using the M1-M2 metrics, it can be concluded that the damage is much more prominent in the GES2N-Max-Np, GES2N-Mean-Np, GES2N-Max-Nf, and GES2N-Mean-Nf's filtered signals than the raw signals results and therefore the methods perform well.

Similarly to the previous section, the damage in measurement $2$ is the most challenging to detect with only the GES2N-Max-Np and the GES2N-Mean-Np enhancing the gear damage components at $1.0$ shaft orders and its harmonics. The ACYCBD can only enhance the gear damage component in measurement $3$ and has similar metrics compared to the GES2N-Max-Np and GES2N-Mean-Np objective functions. The CYCBD, MOMEDA, Spectral Negentropy and the $L_2$/$L_1$ norm could not enhance the damaged gear components for any of the measurements. MOMEDA has, again, the lowest variance (highest M4) of the different methods. Of the GES2N-Max-Np, GES2N-Mean-Np, GES2N-Max-Nf, and GES2N-Mean-Nf objective functions, the GES2N-Max-Np, GES2N-Mean-Np generally have the lowest variance, (i.e., highest M4) when damage is detected, which further highlights that these objective functions perform the best. 

\subsection{Primary sensitivity study}
\label{sec:ExpData:Dataset:SensitivityAnalysis}

For the proposed approach, numerous selections are required, including the filter length, the filter initialisation, the targeted cyclic order, the number of harmonics $N_h$, the cyclic order bands used to estimate the signal and noise components respectively (i.e., cyclic order range and bandwidth), the cyclic order resolution, the weighting of the cyclic orders and the cyclic order bands. Furthermore, the method's sensitivity to these selections should be quantified over multiple measurements. Therefore, an extensive sensitivity study is needed to investigate all parameters. However, this requires a dedicated study and will be done in future work. In this section, a limited but primary sensitivity study is performed of the two best-performing methods, namely, the GES2N-Max-Np and the GES2N-Mean-Np, to the filter length $N_w$, the cyclic order resolution of the SES $\Delta \alpha$ and the cyclic order bands' width $\Delta \alpha_{b}$.

\begin{figure}[h]
	\centering
	\begin{subfigure}{0.32221\linewidth}
		\centering
		\caption{}
		\includegraphics[width=0.97\linewidth]{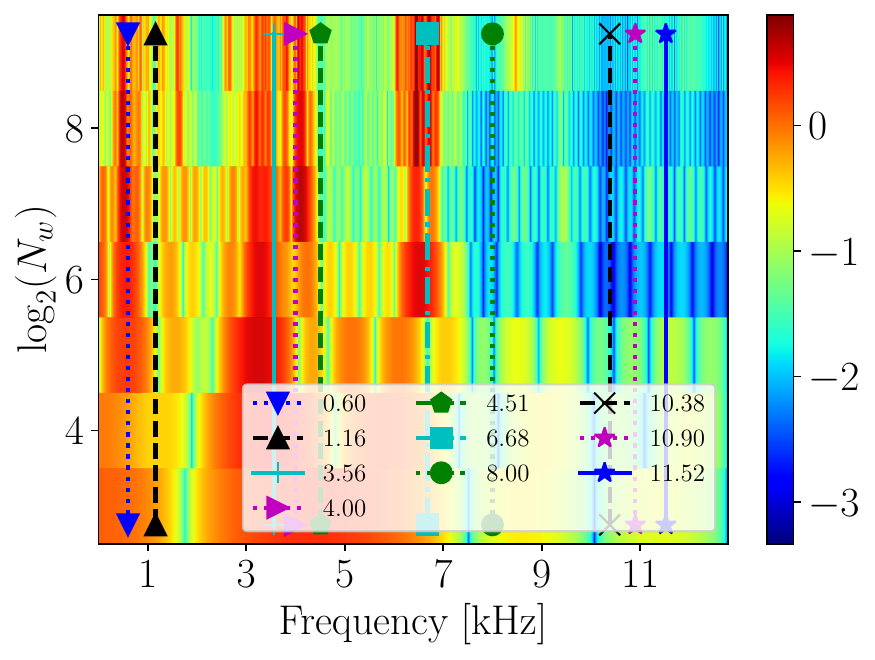}
		\label{fig:Results:Exp:Sensitivity:Exp1D:SNR-Max-Np:FRF-vs-Nw:B0p1_FF0}
	\end{subfigure}	
	\begin{subfigure}{0.32221\linewidth}
		\centering
		\caption{}
		\includegraphics[width=0.97\linewidth]{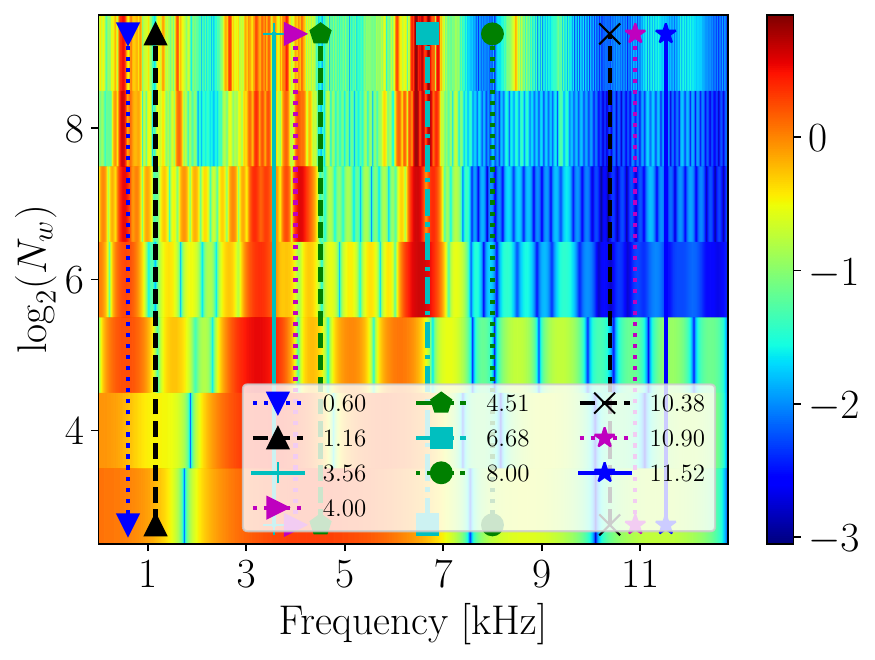}
		\label{fig:Results:Exp:Sensitivity:Exp1D:SNR-Max-Np:FRF-vs-Nw:B0p1_FF1}
	\end{subfigure}
	\begin{subfigure}{0.32221\linewidth}
		\centering
		\caption{}
		\includegraphics[width=0.97\linewidth]{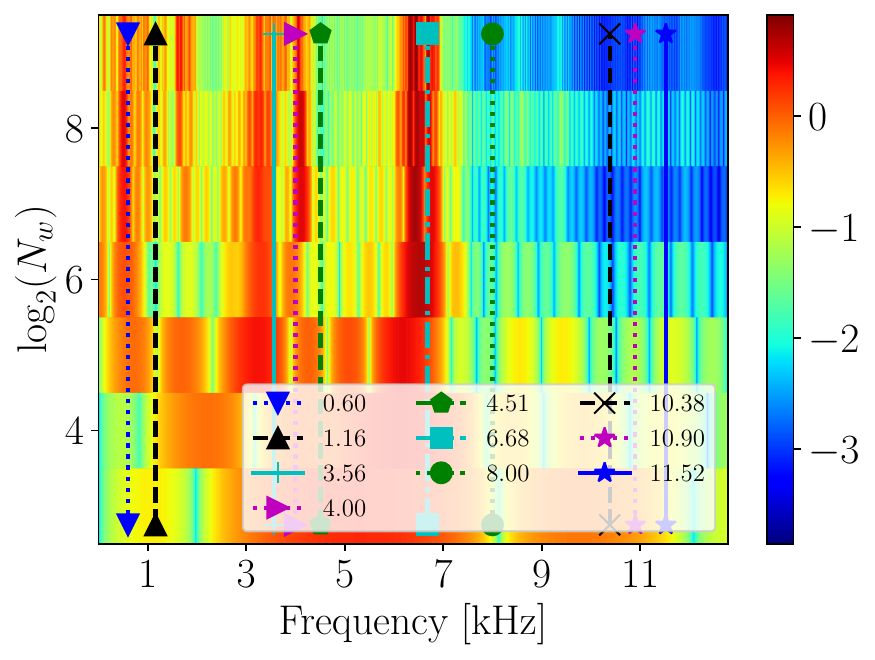}
		\label{fig:Results:Exp:Sensitivity:Exp1D:SNR-Max-Np:FRF-vs-Nw:B0p62_FF1}
	\end{subfigure}
	\begin{subfigure}{0.32221\linewidth}
		\centering
		\caption{}
		\includegraphics[width=0.97\linewidth]{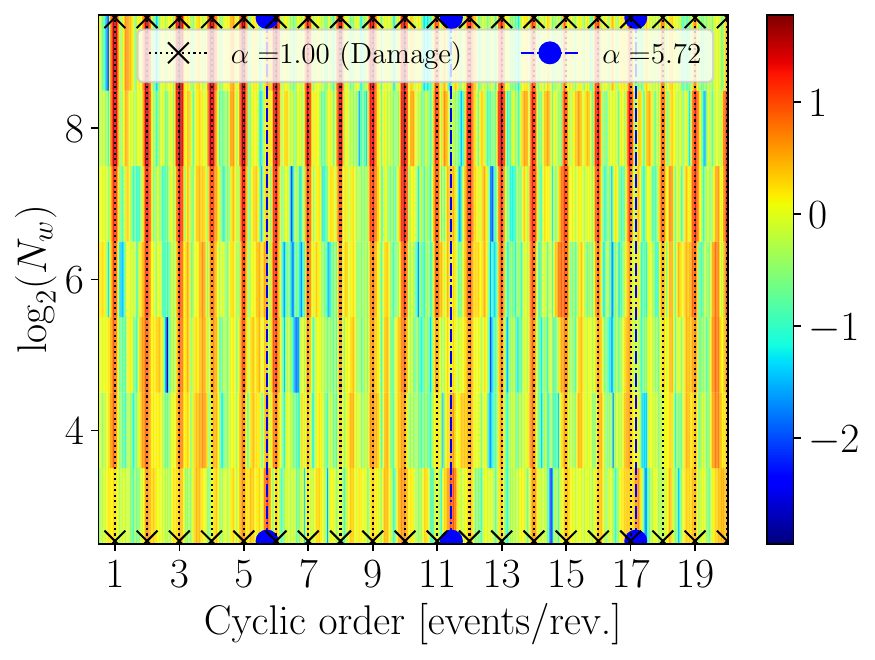}
		\label{fig:Results:Exp:Sensitivity:Exp1D:SNR-Max-Np:SES-vs-Nw:B0p1_FF0}
	\end{subfigure}	
	\begin{subfigure}{0.32221\linewidth}
		\centering
		\caption{}
		\includegraphics[width=0.97\linewidth]{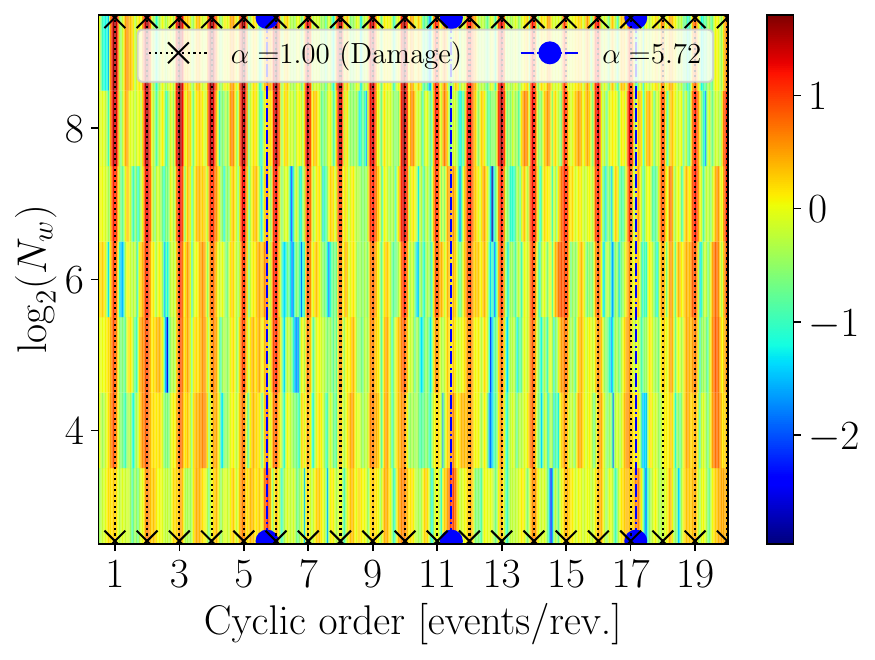}
		\label{fig:Results:Exp:Sensitivity:Exp1D:SNR-Max-Np:SES-vs-Nw:B0p1_FF1}
	\end{subfigure}
	\begin{subfigure}{0.32221\linewidth}
		\centering
		\caption{}
		\includegraphics[width=0.97\linewidth]{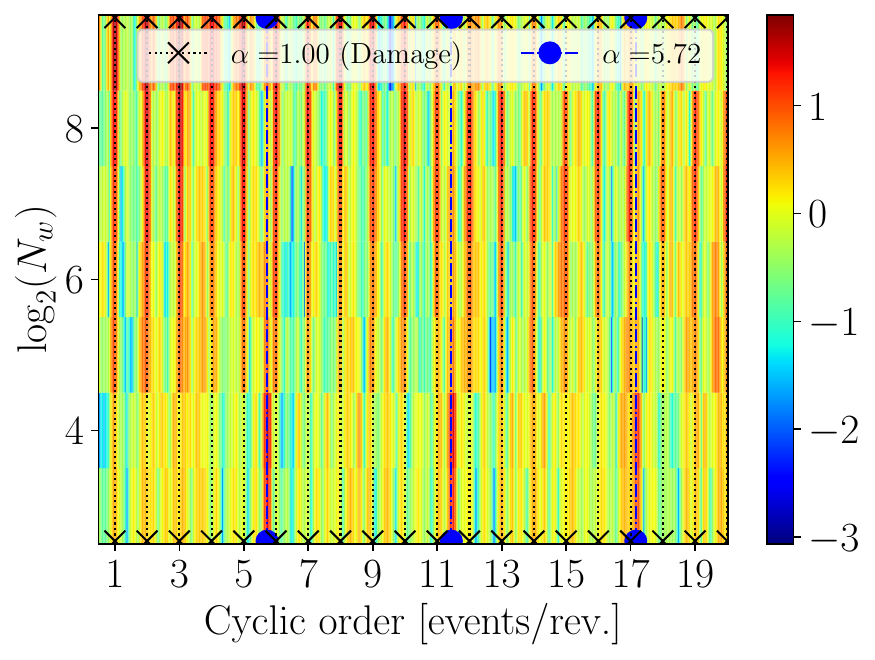}
		\label{fig:Results:Exp:Sensitivity:Exp1D:SNR-Max-Np:SES-vs-Nw:B0p62_FF1}
	\end{subfigure}
	\caption{The performance of the GES2N-Max-Np objective function using different parameters for the cyclic order band's width $\Delta \alpha_{b}$  and the resolution of the cyclic orders $\Delta \alpha$ as a window length $N_{w}$. The top row shows the frequency response of the optimised filtering coefficients, and the bottom row shows the SES of the filtered signal. The following settings are shown: \ref{fig:Results:Exp:Sensitivity:Exp1D:SNR-Max-Np:FRF-vs-Nw:B0p1_FF0} \& \ref{fig:Results:Exp:Sensitivity:Exp1D:SNR-Max-Np:SES-vs-Nw:B0p1_FF0}: $\Delta \alpha_{a} = 0.1, \Delta \alpha = \Delta \alpha_{OT}$; \ref{fig:Results:Exp:Sensitivity:Exp1D:SNR-Max-Np:FRF-vs-Nw:B0p1_FF1} \&  \ref{fig:Results:Exp:Sensitivity:Exp1D:SNR-Max-Np:SES-vs-Nw:B0p1_FF1}: $\Delta \alpha_{a} = 0.1, \Delta \alpha = 0.025$; \ref{fig:Results:Exp:Sensitivity:Exp1D:SNR-Max-Np:FRF-vs-Nw:B0p62_FF1} \& \ref{fig:Results:Exp:Sensitivity:Exp1D:SNR-Max-Np:SES-vs-Nw:B0p62_FF1}: $\Delta \alpha_{a} = 0.62, \Delta \alpha = 0.025$.}
	\label{fig:Results:Exp:Sensitivity:Exp1D:SNR-Max-Np:FRF-vs-Nw}
\end{figure}

\begin{figure}[h]
	\centering
	\begin{subfigure}{0.32221\linewidth}
		\centering
		\caption{}
		\includegraphics[width=0.97\linewidth]{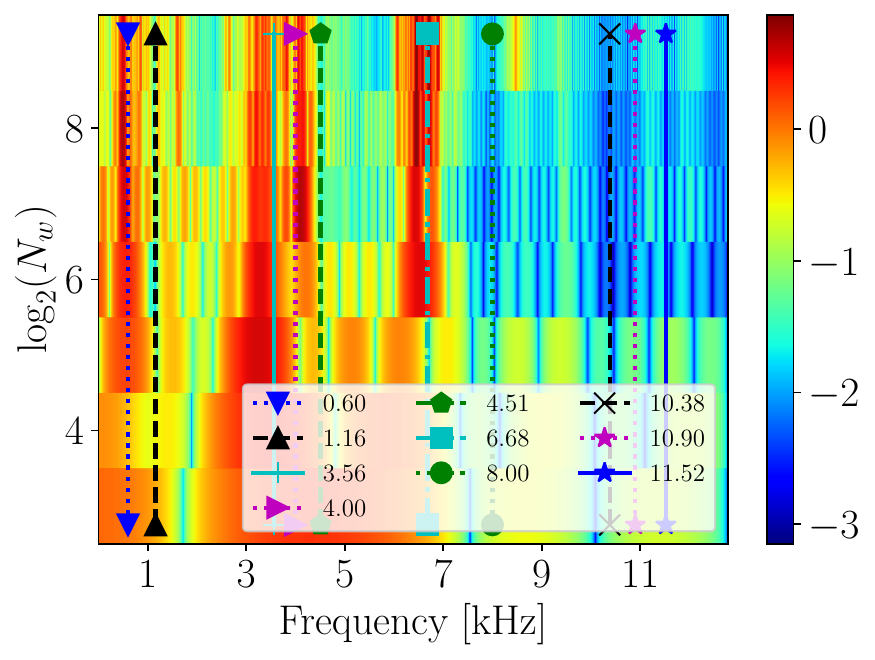}
		\label{fig:Results:Exp:Sensitivity:Exp1D:SNR-Mean-Np:FRF-vs-Nw:B0p1_FF0}
	\end{subfigure}	
	\begin{subfigure}{0.32221\linewidth}
		\centering
		\caption{}
		\includegraphics[width=0.97\linewidth]{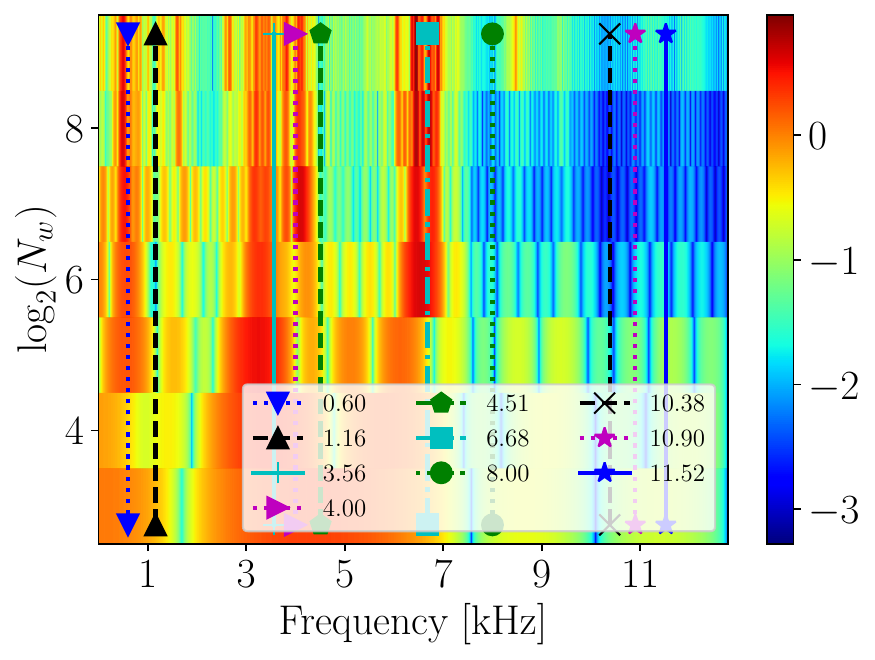}
		\label{fig:Results:Exp:Sensitivity:Exp1D:SNR-Mean-Np:FRF-vs-Nw:B0p1_FF1}
	\end{subfigure}
	\begin{subfigure}{0.32221\linewidth}
		\centering
		\caption{}
		\includegraphics[width=0.97\linewidth]{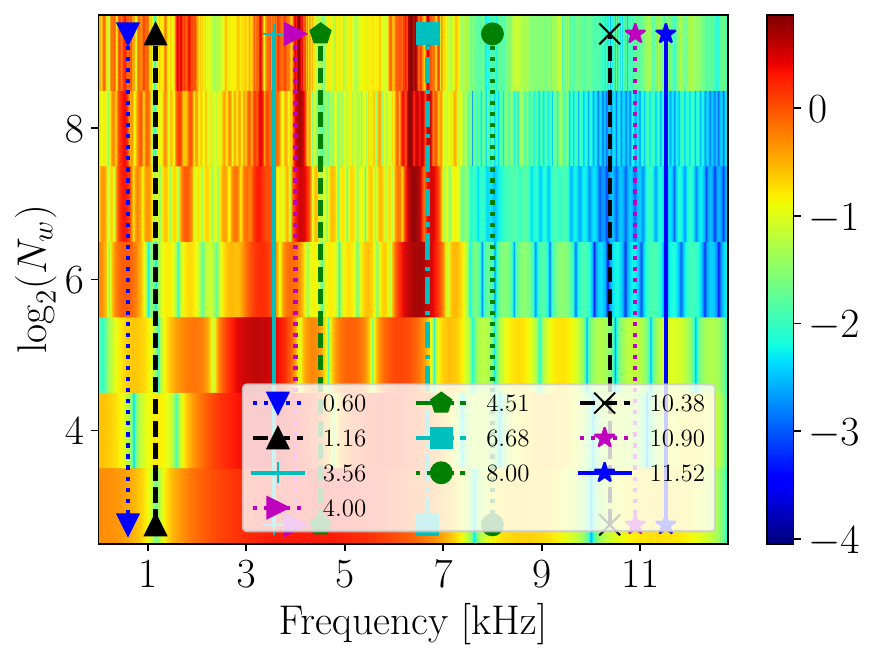}
		\label{fig:Results:Exp:Sensitivity:Exp1D:SNR-Mean-Np:FRF-vs-Nw:B0p62_FF1}
	\end{subfigure}
	\begin{subfigure}{0.32221\linewidth}
		\centering
		\caption{}
		\includegraphics[width=0.97\linewidth]{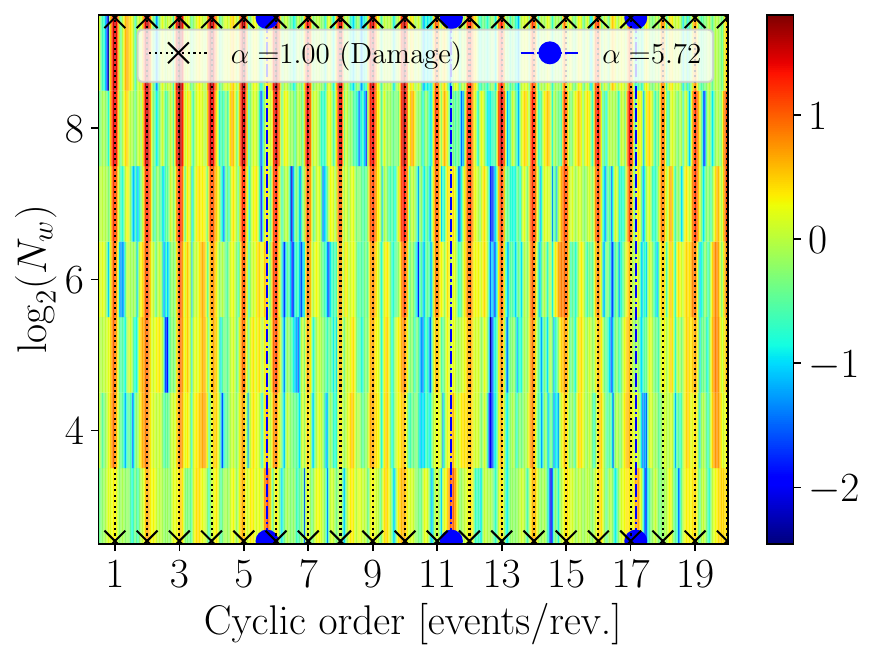}
		\label{fig:Results:Exp:Sensitivity:Exp1D:SNR-Mean-Np:SES-vs-Nw:B0p1_FF0}
	\end{subfigure}	
	\begin{subfigure}{0.32221\linewidth}
		\centering
		\caption{}
		\includegraphics[width=0.97\linewidth]{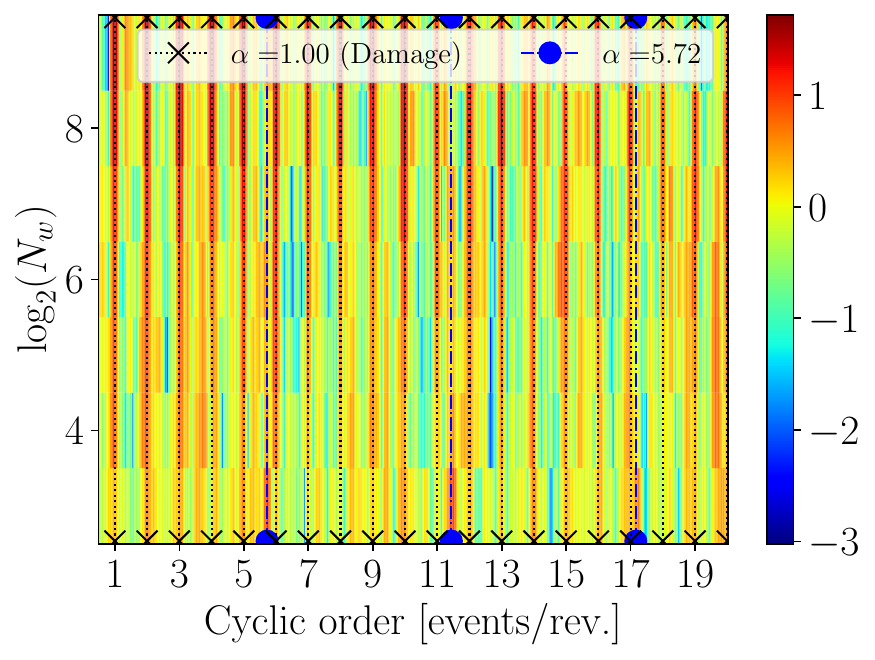}
		\label{fig:Results:Exp:Sensitivity:Exp1D:SNR-Mean-Np:SES-vs-Nw:B0p1_FF1}
	\end{subfigure}
	\begin{subfigure}{0.32221\linewidth}
		\centering
		\caption{}
		\includegraphics[width=0.97\linewidth]{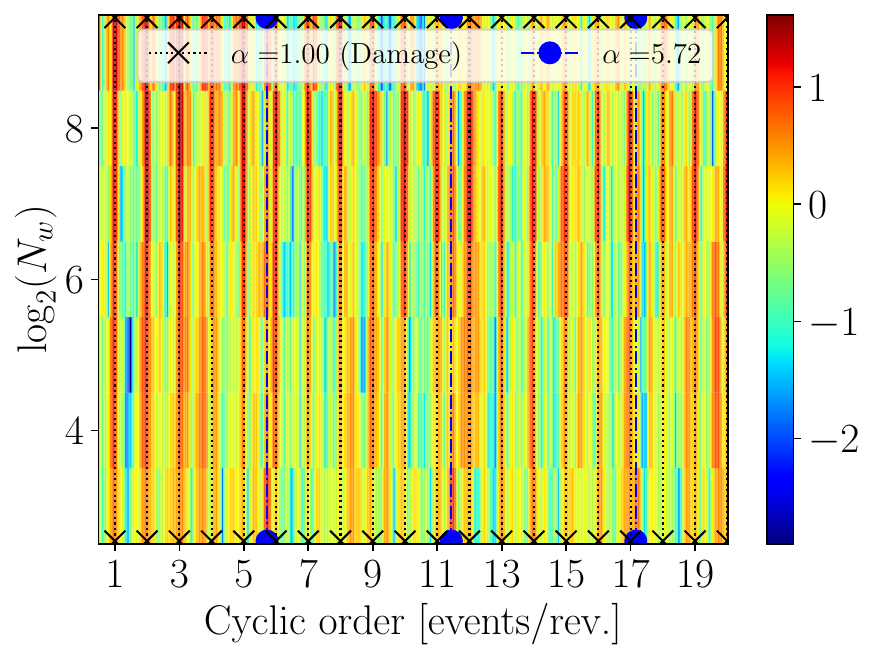}
		\label{fig:Results:Exp:Sensitivity:Exp1D:SNR-Mean-Np:SES-vs-Nw:B0p62_FF1}
	\end{subfigure}
	\caption{The performance of the GES2N-Mean-Np objective function using different parameters for the cyclic order band's width $\Delta \alpha_{b}$  and the resolution of the cyclic orders $\Delta \alpha$ as a window length $N_{w}$. The top row shows the frequency response of the optimised filtering coefficients, and the bottom row shows the SES of the filtered signal. The following settings are shown: \ref{fig:Results:Exp:Sensitivity:Exp1D:SNR-Max-Np:FRF-vs-Nw:B0p1_FF0} \& \ref{fig:Results:Exp:Sensitivity:Exp1D:SNR-Max-Np:SES-vs-Nw:B0p1_FF0}: $\Delta \alpha_{a} = 0.1, \Delta \alpha = \Delta \alpha_{OT}$; \ref{fig:Results:Exp:Sensitivity:Exp1D:SNR-Max-Np:FRF-vs-Nw:B0p1_FF1} \&  \ref{fig:Results:Exp:Sensitivity:Exp1D:SNR-Max-Np:SES-vs-Nw:B0p1_FF1}: $\Delta \alpha_{a} = 0.1, \Delta \alpha = 0.025$; \ref{fig:Results:Exp:Sensitivity:Exp1D:SNR-Max-Np:FRF-vs-Nw:B0p62_FF1} \& \ref{fig:Results:Exp:Sensitivity:Exp1D:SNR-Max-Np:SES-vs-Nw:B0p62_FF1}: $\Delta \alpha_{a} = 0.62, \Delta \alpha = 0.025$.}
	\label{fig:Results:Exp:Sensitivity:Exp1D:SNR-Mean-Np:FRF-vs-Nw}
\end{figure}

In Figure \ref{fig:Results:Exp:Sensitivity:Exp1D:SNR-Max-Np:FRF-vs-Nw} and \ref{fig:Results:Exp:Sensitivity:Exp1D:SNR-Mean-Np:FRF-vs-Nw}, the filter coefficients' frequency response and the SESa of the filtered signal are shown for measurement $2$ of the localised gear damage dataset treated in Section \ref{sec:ExpData:Dataset:LGD}. The results of the GES2N-Max-Np and the GES2N-Mean-Np are included in Figures \ref{fig:Results:Exp:Sensitivity:Exp1D:SNR-Max-Np:FRF-vs-Nw} and \ref{fig:Results:Exp:Sensitivity:Exp1D:SNR-Mean-Np:FRF-vs-Nw} respectively. Two cyclic order resolutions and cyclic order bands' bandwidth are considered: The cyclic order resolution of the SES is either equal to the cyclic order resolution of the order tracked signal's SES $\Delta \alpha = \Delta \alpha_{OT}$ or it is equal to $\Delta \alpha = 0.025 < \Delta \alpha_{OT} $. Furthermore, the cyclic order bands' width is either equal to $0.1$ (i.e., localised closely to the targeted component) or $\Delta \alpha_{b} = 0.62$ such that the extraneous component at $5.72$ shaft orders is included in the numerator's cyclic order bands. 

Figure \ref{fig:Results:Exp:Sensitivity:Exp1D:SNR-Max-Np:FRF-vs-Nw} shows that the GES2N-Max-Np performs better with the longer window lengths than the shorter window lengths. Figure \ref{fig:Results:Exp:Sensitivity:Exp1D:SNR-Max-Np:FRF-vs-Nw:B0p1_FF0} shows that if the window length is short, the filter's frequency response has limited flexibility and therefore cannot effectively enhance the narrow frequency bands and attenuate the extraneous frequency bands. The SES corroborates this in Figure \ref{fig:Results:Exp:Sensitivity:Exp1D:SNR-Max-Np:SES-vs-Nw:B0p1_FF0}, which contains extraneous amplitudes and the targeted amplitudes are not dominant. If the window length is longer, the filter's frequency response has sufficient flexibility to enhance the narrow frequency bands as shown in Figure \ref{fig:Results:Exp:Sensitivity:Exp1D:SNR-Max-Np:FRF-vs-Nw:B0p1_FF0}, which results in fault information rich SESa in Figure \ref{fig:Results:Exp:Sensitivity:Exp1D:SNR-Max-Np:SES-vs-Nw:B0p1_FF0} at the expensive of increased computational cost. Comparing the results in Figures \ref{fig:Results:Exp:Sensitivity:Exp1D:SNR-Max-Np:FRF-vs-Nw:B0p1_FF0},  \ref{fig:Results:Exp:Sensitivity:Exp1D:SNR-Max-Np:FRF-vs-Nw:B0p1_FF1}, \ref{fig:Results:Exp:Sensitivity:Exp1D:SNR-Max-Np:SES-vs-Nw:B0p1_FF0}, and \ref{fig:Results:Exp:Sensitivity:Exp1D:SNR-Max-Np:SES-vs-Nw:B0p1_FF1}, the order tracked signal's cyclic order resolution is sufficient to enhance the gear damage components effectively. 

The cyclic bands' width has a more observable impact on the results in Figure \ref{fig:Results:Exp:Sensitivity:Exp1D:SNR-Max-Np:FRF-vs-Nw:B0p62_FF1} and \ref{fig:Results:Exp:Sensitivity:Exp1D:SNR-Max-Np:SES-vs-Nw:B0p62_FF1}. If the bands' width is too large, the extraneous impulse at $5.72$ shaft order is in the band used in the numerator. Hence, if the maximum amplitude in the spectrum is used in the objective function and the filter length is short, the optimal filter coefficients maximise the extraneous component. However, if the filter length is sufficiently long, the gear damage components and the $5.72$ shaft order components manifest in the SES. This was not investigated, but if the cyclic bands' width is too narrow, it can impede detection if the analytical cyclic order $\alpha_c$ differs from the actual cyclic order in the signal and is not within the targeted band.

The results of the GES2N-Mean-Np are shown in Figure \ref{fig:Results:Exp:Sensitivity:Exp1D:SNR-Mean-Np:FRF-vs-Nw}, with similar observations made relative to the GES2N-Max-Np regarding the filter length and the cyclic order resolution in Figures \ref{fig:Results:Exp:Sensitivity:Exp1D:SNR-Mean-Np:FRF-vs-Nw:B0p1_FF0}, \ref{fig:Results:Exp:Sensitivity:Exp1D:SNR-Mean-Np:SES-vs-Nw:B0p1_FF0}, \ref{fig:Results:Exp:Sensitivity:Exp1D:SNR-Mean-Np:FRF-vs-Nw:B0p1_FF1} and \ref{fig:Results:Exp:Sensitivity:Exp1D:SNR-Mean-Np:SES-vs-Nw:B0p1_FF1}. However, its performance differs from the GES2N-Max-Np when using broad cyclic order bands. Since the GES2N-Mean-Np calculates the mean amplitude in the numerator, the extraneous impulse at $5.72$ orders and the $6$th harmonic of the damage component at $6.0$ orders contribute to the objective function, and, therefore, the damage information is also enhanced. This is seen in Figure \ref{fig:Results:Exp:Sensitivity:Exp1D:SNR-Mean-Np:SES-vs-Nw:B0p62_FF1} where the gear components are much more enhanced than Figure \ref{fig:Results:Exp:Sensitivity:Exp1D:SNR-Max-Np:SES-vs-Nw:B0p62_FF1} for short window lengths. 

\subsection{Discussion}
\label{sec:ExpData:Dataset:Discussion}
The proposed GESR-Max-Np and GESR-Mean-Np objective functions performed the best of all the methods considered, enhancing the gear damage components in the SES for the different measurements and datasets. This makes sense since the specific instance of the objective calculates the ratio between the signal's amplitudes and the noise floor of the SES. In contrast, the GESR-Max-Nf and GESR-Mean-Nf objective functions often performed much worse in cases where the fault signatures are weak in the signal. This highlights that incorporating the $0$ cyclic order component, i.e., the squared energy of the angle domain signal, can degrade the damage enhancement process when the fault signatures are weak. This is because the squared energy of the angle domain signal can be sensitive to extraneous impulsive components, and there might be low-energy spectral frequency bands with non-stationary components. Furthermore, the energy of the filtered signal or its $0$ cyclic order component contains information about the damaged components, which means that the damaged components also contribute to the denominator of the objective, i.e., some of the damaged information forms part of the noise component and therefore is penalised in these objectives. The results further highlight that the maximum operator performs slightly better than the mean for weak damage enhancement. Still, it can be problematic if there are potentially other more dominant signal components in the cyclic order bands.

The results of the MOMEDA, CYCBD and the ACYCBD corroborate that using the energy of the filtered signal in the denominator adversely affected the damage enhancement; the MOMEDA and CYCBD could not enhance the damaged components while the ACYCBD enhanced the damaged components for a few severe damage cases.  The potential negative impact of including the energy of the filtered signal in the denominator is also discussed in Ref. \cite{zhou2022blind}. 

Lastly, the envelope spectra's $L_2$/$L_1$ and spectral negentropy objective functions could not enhance the gear damage components. Their performance is, however, expected since the dataset contains dominant extraneous impulsive components that manifest as sparse components in the SES, masking the weak gear damage components. It is challenging to define all the targeted cyclic orders beforehand  (e.g., all the component's failure modes' cyclic orders). It is also computationally expensive to design filters for each targeted cyclic order. Hence, this work highlights that other approaches (e.g., multiple source extraction) are needed to utilise blind objectives for filter design, and therefore, further research is needed.

\section{Conclusions}
\label{sec:Conclusion:main}

A generalised envelope spectrum-based signal-to-noise objective is proposed from which various objective functions can be derived. The proposed generalised objective is derived for time-varying speed conditions. The proposed objective derives existing objective functions, such as the second-order cyclostationarity objective, and enables the derivation of new objective functions, such as the GES2N-Max-Np. The performance of four objective functions is compared against CYCBD, ACYCBD, MOMEDA, and $L_2$/$L_1$ of the squared envelope spectrum and the spectral negentropy of the squared envelope spectrum for gear damage detection under time-varying speed conditions.  The squared envelope spectra, the filters' frequency response functions, and metrics extracted from the squared envelope spectra are used for comparison. These results demonstrate that the proposed functions perform better than the other methods in enhancing the gear damage under time-varying operating conditions. This is because the proposed generalised objective quantifies the signal-to-noise ratio in the squared envelope spectrum, thereby capturing the prominence of the fault signatures in the squared envelope spectrum.

For future work, we recommend a more extensive comparison against other objectives, that the objective functions be applied to different damage modes and damaged components (e.g., bearings), that the selection of robust and efficient optimisation settings (e.g., optimisers and initialisation strategies) be automated, an extensive sensitivity study of the hyperparameters and extending the signal-to-noise ratio objective to more general settings.

\section*{Acknowledgements}

S. Schmidt gratefully acknowledges the Research and Development Programme of the University of Pretoria for supporting the research.
K.C. Gryllias gratefully acknowledges the financial support provided by the FWO - Fonds Wetenschappelijk Onderzoek - Vlaanderen under the project G0A3123N.

\bibliographystyle{elsarticle-num} 
\bibliography{my_bib}

\appendix 

\section{Code}
\label{app:GitLabCode}

The following repository contains an implementation of the proposed objective function: \url{https://gitlab.com/vspa/snrocl}.

\section{Gradient of the objective function}
\label{app:DerivationGradient}

The objective function, defined in Equation \eqref{eqn:Method:Formulation2}, is repeated here for the reader's convenience: 
\begin{equation}
	\label{eqn:Method:MainObjective_Log_DEFINITION2}
	\ln\psi\left(
	\boldsymbol{g}; \boldsymbol{x}
	\right) = \ln\left(
	\boldsymbol{w}_{s}^\textrm{T}\boldsymbol{C}_{s}\boldsymbol{b}
	\right) - \ln\left(
	\boldsymbol{w}_{n}^\textrm{T}\boldsymbol{C}_{n}\boldsymbol{b}
	\right).
\end{equation}
The analytical gradient of this objective function with respect to the design variables, i.e., the filter coefficients, is defined as follows:
\begin{equation}
\label{eqn:Method:MainObjective_Gradient_Summarised}
	\frac{d}{d\boldsymbol{h}} \ln\psi\left(
	\boldsymbol{g}; \boldsymbol{x}
	\right) = \frac{d \boldsymbol{g}}{d \boldsymbol{h}} \left(
\frac{d}{d\boldsymbol{g}} \ln\psi\left(
\boldsymbol{g}; \boldsymbol{x}
\right) 
	\right),
\end{equation}
where $\frac{d}{d\boldsymbol{h}} \ln\psi\left(
	\boldsymbol{g}; \boldsymbol{x}
	\right):\mathbb{R}^{D \times 1} \mapsto \mathbb{R}^{D \times 1}$. The first gradient component in Equation \eqref{eqn:Method:MainObjective_Gradient_Summarised} is given by
\begin{equation}
    \frac{d \boldsymbol{g}}{d \boldsymbol{h}} = \frac{||\boldsymbol{h}||_{2}^2 \cdot \boldsymbol{I}- \boldsymbol{h}\boldsymbol{h}^\textrm{T}}{||\boldsymbol{h}||_{2}^3},
\end{equation}
with $\frac{d \boldsymbol{g}}{d \boldsymbol{h}}: \mathbb{R}^{D \times 1} \mapsto \mathbb{R}^{D \times D}$, and the identity matrix denoted $\boldsymbol{I} \in \mathbb{R}^{D \times D}$.
The second gradient component in Equation \eqref{eqn:Method:MainObjective_Gradient_Summarised} is given by
\begin{equation}
        \label{eqn:Method:MainObjective_GradientLog_Full}
	\frac{d}{d\boldsymbol{g}} \ln\psi\left(
	\boldsymbol{g}; \boldsymbol{x}
	\right) = \frac{d\boldsymbol{b}}{d\boldsymbol{g}}\left(\frac{\boldsymbol{w}_{s}^\textrm{T}\boldsymbol{C}_{s}}{\boldsymbol{w}_{s}^\textrm{T}\boldsymbol{C}_{s}\boldsymbol{b}}   - \frac{\boldsymbol{w}_{n}^\textrm{T}\boldsymbol{C}_{n} }{\boldsymbol{w}_{n}^\textrm{T}\boldsymbol{C}_{n}\boldsymbol{b}} \right)^\textrm{T}, \text{ with } \frac{d}{d\boldsymbol{g}} \ln\psi\left(
	\boldsymbol{g}; \boldsymbol{x}
	\right):\mathbb{R}^{D \times 1} \mapsto \mathbb{R}^{D \times 1},
\end{equation}
if $\boldsymbol{w}_i$ and $\boldsymbol{C}_i$ are not dependent on the filter coefficients, i.e., $\frac{d \boldsymbol{w}_i}{d \boldsymbol{g}} = \boldsymbol{0}$ and $\frac{d \boldsymbol{C}_i}{d \boldsymbol{g}} = \boldsymbol{0}$ for the numerator and the denominator. Hence, for cases where the vectors and matrices are functions of the data, the sensitivity in Equation \eqref{eqn:Method:MainObjective_GradientLog_Full} is incomplete. The gradient $\frac{d \boldsymbol{b}}{d \boldsymbol{g}}$ in Equation \eqref{eqn:Method:MainObjective_GradientLog_Full} is defined as follows:
\begin{align}
	\frac{d \boldsymbol{b}}{d \boldsymbol{g}} &= \frac{d}{d \boldsymbol{g}}(\boldsymbol{V} \left(
	\boldsymbol{y} \odot \boldsymbol{y}
	\right))^{\boldsymbol{*}} \odot \left(\boldsymbol{V} \left(
	\boldsymbol{y} \odot \boldsymbol{y}
	\right)\right), \\
	\frac{d \boldsymbol{b}}{d \boldsymbol{g}} &=    4  \cdot \left[
 \left(\boldsymbol{V} \left(
	\boldsymbol{y} \odot \boldsymbol{y}
	\right)\right) \odot \left(\boldsymbol{V}^{\boldsymbol{*}} \left(
	\boldsymbol{y} \odot \boldsymbol{X}
	\right)\right)
 \right]^\textrm{T},
\end{align}
where $\frac{d \boldsymbol{b}}{d \boldsymbol{g}} 
 \in \mathbb{R}^{D \times N_f}$, $\boldsymbol{y} = \boldsymbol{X} \boldsymbol{g}$, $\frac{d \boldsymbol{y}}{d\boldsymbol{g}} = \boldsymbol{X}^\textrm{T}$ and $\boldsymbol{y} \odot \boldsymbol{X}$ denotes the Hadamard product between $\boldsymbol{y}$ and each column of $\boldsymbol{X}$, i.e., $\boldsymbol{y} \odot \boldsymbol{X} \in \mathbb{R}^{L_y \times D}$.

\end{document}